\documentclass[12pt,a4paper, fleqn]{article}

\RequirePackage[l2tabu, orthodox]{nag}

\usepackage{mjheppub}

\usepackage{graphicx}

\usepackage{tikz}

\usepackage{siunitx}

\usepackage[T1]{fontenc}
\usepackage[utf8]{inputenc}

\usepackage{pgfplots}
\pgfplotsset{compat=newest}

\newcommand*{\colorboxed}{}
\def\colorboxed#1#{%
  \colorboxedAux{#1}%
}
\newcommand*{\colorboxedAux}[3]{%
  \begingroup
    \colorlet{cb@saved}{.}%
    \color#1{#2}%
    \boxed{%
      \color{cb@saved}%
      #3%
    }%
  \endgroup
}

\title{The analytic bootstrap equations \\ of non-diagonal two-dimensional CFT}

\author{Santiago Migliaccio and Sylvain Ribault}

\affiliation{\vspace{2mm}
Institut de physique th\'eorique, Universit\'e Paris Saclay, CEA, CNRS, F-91191 Gif-sur-Yvette}

\emailAdd{santiago.migliaccio@ipht.fr, sylvain.ribault@ipht.fr}

\abstract{Under the assumption that degenerate fields exist, diagonal CFTs such as Liouville theory can be solved analytically using the conformal bootstrap method. Here we generalize this approach to non-diagonal CFTs, i.e. CFTs whose primary fields have nonzero conformal spins. Assuming generic values of the central charge, we find that the non-diagonal sector of the spectrum must be parametrized by two integer numbers. We then derive and solve the equations that determine how three- and four-point structure constants depend on these numbers. In order to test these results, we numerically check crossing symmetry of a class of four-point functions in a non-rational limit of D-series minimal models.
The simplest four-point functions in this class were previously argued to describe connectivities of clusters in the critical Potts model.
}

\begin{document}

\maketitle

\flushbottom

\section{Introduction and summary}

The conformal bootstrap method is arguably the simplest way of exactly computing correlation functions in diagonal CFTs such as Liouville theory and minimal models. 
The idea is to constrain generic three-point structure constants by studying four-point functions that involve degenerate fields. 
This is possible provided that degenerate fields exist, i.e. that OPEs and correlation functions involving degenerate fields make sense and obey axioms such as OPE associativity. 
In this sense, degenerate fields can exist without ever appearing in OPEs of other fields: this is known to happen in Liouville theory. (See \cite{rib16} for a brief review.)

In this article we will investigate whether the same method can be applied to the case of non-diagonal CFTs. There is ample motivation for investigating these theories: for example, the Potts model at criticality is expected to display non-diagonal features. (See \cite{ei15} and references therein.)
However, we will not attempt to solve any specific model, but rather compute correlation functions that obey conformal bootstrap equations. 
We will make three main assumptions which will allow us to derive such equations, and which are known to hold in consistent conformal field theories such as Liouville theory and generalized minimal models \cite{Ribault:2014hia}. We assume
\begin{enumerate}
 \item that two independent degenerate fields exist,
 \item that correlation functions are single-valued, 
 \item that the model depends analytically on the central charge.
\end{enumerate}
As a first consequence of these assumptions, we will show that non-diagonal primary fields are parametrized by two integer numbers. 
(By non-diagonal fields we mean not only spinful fields, but also spinless fields that generate spinful fields when fused with degenerate fields.) 
Each degenerate field will be responsible for shifting the value of one of the integer numbers, and the resulting bootstrap equations will therefore determine the correlation functions of our non-diagonal fields.

Some steps in this direction were previously taken by Estienne and Ikhlef \cite{ei15}, who found the striking result that three-point structure constants of certain spinful fields were geometric means of Liouville three-point structure constants. 
(See also \cite{hj13} for a similar relation in a more complicated CFT.)
Schematically,
\begin{align}
 C(\Delta_i,\bar \Delta_i) = \sqrt{C_\text{L}(\Delta_i) C_\text{L}(\bar\Delta_i)}\ ,
 \label{eq:csqrt}
\end{align}
where $\Delta_i$ and $\bar \Delta_i$ are left- and right-moving conformal dimensions respectively, so that a primary field is spinful if $\Delta_i\neq \bar\Delta_i$ and spinless if $\Delta_i= \bar\Delta_i$.
The structure constants that we will compute do obey a geometric mean relation, 
where we will resolve the sign ambiguity and show that the square root is compatible with an analytic dependence on the central charge.
In Appendix \ref{app} we will actually show that the geometric mean relation is a universal feature of non-diagonal CFTs under certain assumptions.

The bootstrap equations that determine the structure constants are derived from four-point functions that involve degenerate fields, 
but they do not imply that more general four-point functions are crossing-symmetric. 
In order to show that, we also need to determine the operator product expansions (OPEs) of the fields. And finding which fields appear in a given OPE is not necessarily easy.
For instance, Liouville theory with a central charge less than one
was shown to exist by the determination of its spectrum and OPEs \cite{rs15} a long time after its structure constants were computed. 

In order to guess plausible OPEs we will take a limit of the D-series Virasoro minimal models, a class of non-diagonal CFTs which have already been solved, but which exist only for discrete values of the central charge. 
Since however these values are dense in the half-line $c\in(-\infty,1)$, taking limits of minimal model OPEs yields sensible OPEs for any $c\in(-\infty,1)$, and actually by analyticity for any central charge in the half-plane
\begin{align}
 \Re c < 13\ .
\end{align}
By combining these OPEs and our analytic structure constants, we can compute four-point functions  of two diagonal fields with arbitrary conformal dimensions, and two non-diagonal fields. We will numerically check that these four-point functions are crossing-symmetric. 

The D-series minimal models, along with their limits and analytic continuations, obey a rule of conservation of diagonality: a correlation function can be nonzero only if it involves an even number of non-diagonal fields. This implies that our four-point functions based on limits of D-series minimal models only involve three-point structure constants with $0$ or $2$ non-diagonal fields. In order to test our results with $1$ or $3$ non-diagonal fields, we will focus on the Ashkin--Teller model, a $c=1$ CFT where diagonality is not conserved. We will show that our analytic results agree with Al. Zamolodchikov's \cite{zam85b} for this model.

\section{Spinful fields and their correlation functions}

\subsection{Global conformal symmetry}

Consider a field theory on the Riemann sphere, with local conformal symmetry. The symmetry algebra is made of two copies of the Virasoro algebra, with the same central charge $c$. 
For the moment, let us focus on global conformal transformations,
\begin{align}
z \rightarrow \frac{az+b}{cz+d} \quad , \quad \text{with}\ \left( \begin{array}{cc} a & b\\c& d \end{array} \right) \in SL_2(\mathbb{C})\ .
\end{align}
Under such transformations, a primary field $V(z)$ with left and right conformal dimensions $\Delta$ and $\bar{\Delta}$ behaves as 
\begin{align}
V(z)\rightarrow (cz+d)^{-2\Delta}(\bar{c}\bar{z}+\bar{d})^{-2\bar{\Delta}}V\left(\frac{az+b}{cz+d}\right)\, .
\end{align}
In particular, a rotation by an angle $\theta$ corresponds to  $\left( \begin{smallmatrix} e^{i\frac{\theta}{2}} & 0\\0& e^{-i\frac{\theta}{2}} \end{smallmatrix}\right)$, and gives
\begin{align}
V(z)\rightarrow e^{-i\theta(\Delta-\bar{\Delta})}V\left(e^{i\theta}z\right)\, .  
\end{align}
In this transformation, there appears the difference between the right and left dimensions, called the conformal spin and denoted as
\begin{align}
S = \Delta - \bar{\Delta} \, . 
\end{align}
In a CFT, correlation functions are invariant under conformal transformations. 
In particular, the invariance of an $n$-point function of primary fields $V_{i}(z_i)$ with spins $S_i$ implies
\begin{align}
\left\langle \prod_{i=1}^{n}V_{i}(z_i) \right\rangle = e^{-i\theta \sum_{j=1}^{n}S_j} \left\langle \prod_{i=1}^{n}V_{i}(e^{i\theta}z_i) \right\rangle\, .\label{rotn}
\end{align}
Assuming that our $n$-point function is single-valued and non-vanishing, the phase should be $1$ for the rotation by $\theta = 2\pi$, which implies
\begin{align}
\sum_{j=1}^{n} S_j \in \mathbb{Z}\, . \label{sumsp}
\end{align}

Let us consider the cases of two-point functions. Global conformal symmetry implies that a two-point function vanishes unless the two fields have the same left and right conformal dimensions. We further assume that there is an orthogonal basis of primary fields, so that two-point functions of elements of this basis take the form
\begin{align}
\Big\langle V_1(z_1)V_2(z_2)\Big\rangle = B_1 \frac{\delta_{12}}{z_{12}^{2\Delta_1}\bar{z}_{12}^{2\bar{\Delta}_1}}\, ,
\label{2pt}
\end{align}
where $B_1 = B(V_1)$ is a $z$-independent factor called the two-point structure constant.
In particular, the two-point function can be nonzero only if $S_1 = S_2$. Combined with the constraint on the total spin eq. \eqref{sumsp}, this implies that the spins $S_1,S_2$ obey
\begin{align}
S \in \frac{1}{2}\mathbb{Z}\, .\label{sZ2}
\end{align}
All our fields will obey this constraint, and will be called bosons if $S \in \mathbb{Z}$, and fermions if $S \in \mathbb{Z} +\frac{1}{2}$. 

In the case of three-point functions, global conformal symmetry implies
\begin{multline}
\Big\langle V_1(z_1) V_2(z_2) V_3(z_3) \Big\rangle = C_{123} \\
\times z_{12}^{\Delta_3-\Delta_2 -\Delta_1}\bar{z}_{12}^{\bar{\Delta}_3-\bar{\Delta}_2 -\bar{\Delta}_1}z_{23}^{\Delta_1-\Delta_2 -\Delta_3}\bar{z}_{23}^{\bar{\Delta}_1-\bar{\Delta}_2 -\bar{\Delta}_3} z_{31}^{\Delta_2-\Delta_3 -\Delta_1} \bar{z}_{31}^{\bar{\Delta}_2-\bar{\Delta}_3 -\bar{\Delta}_1}\, , \label{3p}
\end{multline}
where $C_{123}= C(V_1,V_2,V_3)$ is called the three-point structure constant. 
Since fermions anticommute,  under a permutation $\sigma$ the three-point function should behave as
\begin{align}
\Big\langle V_{\sigma(1)}V_{\sigma(2)}V_{\sigma(3)}\Big\rangle = \eta_{123}(\sigma) \Big\langle V_1 V_2 V_3 \Big\rangle\, ,  
\end{align}
where
\begin{align}
\eta_{123}(\sigma) = \begin{cases}
-1 \quad &\text{if } \sigma \text{ exchanges two fermions,}\\
1 \quad  &\text{else.} \end{cases}\, 
\end{align}
In order to compensate for the behaviour of the $z$-dependent factor of the three-point function \eqref{3p}, the structure  constant $C_{123}$ should therefore satisfy
\begin{align}
\boxed{\frac{C_{\sigma(1)\sigma(2)\sigma(3)}}{C_{123}} = \eta_{123}(\sigma )\,\text{sgn}(\sigma)^{S_1 + S_2 + S_3}} \, .\label{permC}
\end{align}

\subsection{Diagonal and non-diagonal fields}\label{sec:dnd}

In order to determine the three-point structure constants, we will use constraints coming from four-point functions that involve degenerate fields. 
Before writing fusion rules involving degenerate fields, let us introduce notations that make them simpler. 
We have the alternative notation $\beta$ for the central charge $c$,
\begin{align}
c = 1 - 6\left(\beta-\frac{1}{\beta}\right)^2\, ,
\end{align}
and we introduce the momentum $P$ instead of the conformal dimension $\Delta$,
\begin{align}
\Delta = \frac{c-1}{24} + P^2\, .
\end{align}
In terms of momentums, the spin is 
\begin{align}
S = P^2 - \bar{P}^2 \,.
\end{align}
We now introduce the diagonal degenerate fields $V_{\langle 2,1 \rangle}$ and $V_{\langle 1,2 \rangle}$. Each one of these fields has vanishing null vectors at level two for both the left- and right-moving Virasoro algebras. Each one of these fields has the same left and right momentums, whose values $P_{(2,1)}$ and $P_{(1,2)}$ are special cases of eq. \eqref{eq:prs}. We now write these momentums, together with the corresponding conformal dimensions:
\begin{align}
P_{(2,1)} = \beta - \frac{1}{2\beta} & \implies 
\Delta_{(2,1)} = -\frac12 + \frac{3\beta^2}{4}  \ , 
\\
P_{(1,2)} = \frac{\beta}{2} - \frac{1}{\beta} &\implies  
\Delta_{(1,2)} = -\frac12 + \frac{3}{4\beta^2}\ .
\end{align}
Writing $V_{P,\bar{P}}$ a primary field with left and right momentums $P$ and $\bar P$, its fusion product with $V_{\langle 2,1 \rangle}$ takes the form
\begin{align}
V_{\langle 2,1 \rangle} \times V_{P,\bar{P}} \subseteq \sum_{\epsilon = \pm,\, \bar{\epsilon} = \pm} V_{P+ \tfrac{\epsilon \beta}{2},\bar{P}+ \tfrac{\bar{\epsilon}\beta}{2}}\, . \label{V214}
\end{align}
This fusion rule follows from the existence of vanishing descendents of $V_{\langle 2,1 \rangle}$. Let us furthermore assume that the fields satisfy the half-integer spin condition \eqref{sZ2}. 
Then the difference of the spins of $V_{P,\bar{P}}$ and $V_{P+ \tfrac{\epsilon \beta}{2},\bar{P}+ \tfrac{\bar{\epsilon}\beta}{2}}$ must be half-integer, which implies 
\begin{align}
P-\sigma \bar{P} \in \frac{1}{2\beta}\mathbb{Z}\, , \label{spinc21}
\end{align}
with $\sigma =\epsilon\bar{\epsilon} \in \{+,-\}$. If this holds for both values of $\sigma$, then it follows that 
\begin{align}
S = \prod_{\pm} (P\pm\bar{P}) \in \frac{1}{4\beta^2}\mathbb{Z}\, .
\end{align}
Assuming that $S\neq 0$ and that the central charge is generic, so that $\beta^2\notin\mathbb{Q}$, this is incompatible with the half-integer spin condition \eqref{sZ2}.
Therefore, eq. \eqref{spinc21} can be satisfied for only one value of $\sigma$, and the terms in the fusion product $V_{\langle 2,1 \rangle} \times V_{P,\bar{P}}$ with the other value of $\sigma$ must, in fact, be absent.
Similarly, the fusion product $V_{\langle 1,2 \rangle} \times V_{P,\bar{P}}$ can have only two terms, and these terms are determined by the value of the sign $\tilde{\sigma}\in\{+,-\}$ such that
\begin{align}
P-\tilde{\sigma} \bar{P} \in \frac{ \beta }{2}\mathbb{Z}\, . \label{spinc12}
\end{align}
We consider $\sigma$ and $\tilde{\sigma}$ as properties of fields, which control their fusion products with degenerate fields:
\begin{align}
V_{\langle 2,1 \rangle} \times V_{P,\bar{P}}^{\sigma, \tilde{\sigma}} = \sum_{\epsilon = \pm} V_{P+\epsilon \tfrac{\beta}{2},\bar{P}+\sigma\epsilon \tfrac{\beta}{2}}^{\sigma, \tilde{\sigma}}\quad ,
\quad
V_{\langle 1,2 \rangle} \times V_{P,\bar{P}}^{\sigma, \tilde{\sigma}} = \sum_{\epsilon = \pm} V_{P-\tfrac{\epsilon}{2\beta},\bar{P}- \tfrac{\tilde{\sigma}\epsilon}{2\beta}}^{\sigma, \tilde{\sigma}}\, .\label{dfus}
\end{align}
The fields on the right-hand sides have the same values of $\sigma,\tilde{\sigma}$ as the field $V_{P,\bar{P}}^{\sigma, \tilde{\sigma}}$, because their momentums satisfy the same relations \eqref{spinc21},\eqref{spinc12}. 
Let us now solve these relations and determine the momentums. The solution strongly depends on $\sigma\tilde{\sigma}$, and we will call fields diagonal or non-diagonal depending on this sign. Once $\sigma\tilde{\sigma}$ is fixed, the choice of $\sigma$ makes no difference: flipping $\sigma$ is equivalent to performing the reflection $(P, \bar P) \to (P, -\bar P)$, which leaves the conformal dimensions $(\Delta,\bar\Delta)$ invariant. 
In what follows we set $\sigma=+$ without loss of generality.
\begin{itemize}
\item \textbf{Diagonal fields $\tilde{\sigma} = \sigma$}

Still assuming $\beta^2\notin\mathbb{Q}$, diagonal fields must have $P - \sigma \bar{P}\in \frac{ \beta }{2}\mathbb{Z} \cap \frac{1}{2\beta}\mathbb{Z} = \{0\} $, and therefore $\Delta=\bar{\Delta}$. Introducing the notation
\begin{align}
\boxed{V^{D}_P = V^{+, +}_{P,P}} \, ,
\end{align}
the fusion products with degenerate fields read
\begin{align}
V_{\langle 2,1 \rangle} \times V_{P}^D = \sum_{\epsilon = \pm} V_{P+\frac{\epsilon \beta}{2}}^{D}\quad ,\quad V_{\langle 1,2 \rangle} \times V_{P}^{D} = \sum_{\epsilon = \pm} V_{P-\frac{\epsilon}{2\beta}}^{D}\, .
\end{align}

\item \textbf{Non-diagonal fields $\tilde{\sigma} = -\sigma$}

Keeping in mind the half-integer spin condition \eqref{sZ2}, the momentums of non-diagonal fields are of the type
\begin{align}
\left\{\begin{array}{l}
P=P_{( r,s )}\, , \\ \bar{P}= \sigma P_{( r,-s )}\, , \end{array}\right. \quad 
\text{with}\quad  r,s,rs \in \frac{1}{2}\mathbb{Z}\, ,
\label{eq:rsval}
\end{align}
where we introduced the notation
\begin{align}
P_{( r,s)}=\frac{1}{2}\left(r\beta - \frac{s}{\beta}\right)\, .
\label{eq:prs}
\end{align}
The spin of a non-diagonal field is $S = -rs$. Introducing the notation
\begin{align}
\boxed{V^N_{( r,s )} = V^{+,-}_{P_{( r,s)},P_{( r,-s)}}} \, ,
\end{align}
the fusion products with degenerate fields read
\begin{align}
V_{\langle 2,1 \rangle}\times V_{( r,s)}^N = \sum_{\epsilon = \pm}  V_{( r+\epsilon,s)}^{N}\quad ,\quad V_{\langle 1,2 \rangle} \times V_{(r,s)}^{N} = \sum_{\epsilon = \pm}  V_{( r,s+\epsilon)}^N\, .
\end{align}
\end{itemize}
We emphasize that it is not enough to know the momentums of a field in order to identify it as diagonal or non-diagonal. 
The non-diagonal field $V^N_{(r,0)}$ has spin zero, and the same left and right momentums as the diagonal field $V^{D}_{P_{(r,0)}}$. These two fields are distinguished by their fusion products with degenerate fields, equivalently by their values of $\sigma$ and $\tilde{\sigma}$. In particular, the fusion product 
\begin{align}
V_{\langle 1,2 \rangle}\times V^{N}_{(r,0)} = \sum_{\epsilon = \pm} V^{N}_{(r, \epsilon)}\, ,
\end{align}
produces spinful fields. Notice also that linear combinations of $V^D_{P_{(r,0)}}$ and $V^N_{(r,0)}$ are primary fields whose fusion products with $V_{\langle 2,1\rangle}$ and $V_{\langle 1,2\rangle}$ can involve four fields, rather than two fields as in the fusion products \eqref{dfus}. And indeed the derivation of our fusion products assumed that the spin $S$ was nonzero, while we are now dealing with spinless fields of momentums $P=\bar P = P_{(r,0)}$. Linear combinations of $V^D_{P_{(r,0)}}$ and $V^N_{(r,0)}$ might not be the only fields whose fusion products with $V_{\langle 2,1\rangle}$ and $V_{\langle 1,2\rangle}$ involve four fields, but we refrain from studying such special cases in more detail, and restrict our attention to our diagonal and non-diagonal fields.

\section{Analytic conformal bootstrap}\label{sec:acb}

In this Section we will derive and solve conformal bootstrap equations for correlation functions of our diagonal and non-diagonal fields. These equations will follow from the assumption that the correlation functions $\left\langle V_{\langle 2,1 \rangle} V_1 V_2 V_3 \right\rangle $  and $\left\langle V_{\langle 1,2 \rangle} V_1 V_2 V_3 \right\rangle $  exist, for  arbitrary diagonal or non-diagonal fields $V_1,V_2,V_3$.

\subsection{Operator product expansions and crossing symmetry}\label{ssec:OPE}

We will write operator product expansions in a schematic notation that omits the dependence on the coordinates $z_i$, and also omits the contributions of descendent fields:
\begin{align}
V_1 V_2 = \sum_{V_3 \in \mathbb{S} } C_{12}^3  V_3\, . \label{gOPE}
\end{align}
Here $\mathbb{S}$ is a set of primary fields that we call the spectrum of the OPE. 
The spectrum of an OPE is a subset of the spectrum of the theory.
Inserting the OPE into the three-point function \eqref{3p}, we find that the OPE coefficient $C_{12}^3$ is related to the two- and three-point structure constants $B_3$ \eqref{2pt} and $C_{123}$ by
\begin{align}
C_{123} = B_3 C_{12}^3 \, .
\end{align}
Let us write the degenerate OPEs that correspond to the fusion rules \eqref{dfus} as 
\begin{align}
V_{\langle 2,1 \rangle} V
= \sum_{\epsilon = \pm} C_{\epsilon}(V) V^\epsilon
\quad ,\quad 
V_{\langle 1,2 \rangle} V
= \sum_{\epsilon = \pm} \tilde{C}_{\epsilon}(V) V^{\tilde{\epsilon}} \, ,\label{dOPE}
\end{align}
where we introduced the notations 
\begin{align}
 V = V_{P,\bar{P}}^{\sigma, \tilde{\sigma}} 
 \quad \implies \quad V^\epsilon = V_{P+\epsilon \tfrac{\beta}{2},\bar{P}+\sigma\epsilon \tfrac{\beta}{2}}^{\sigma, \tilde{\sigma}}\quad ,\quad 
 V^{\tilde{\epsilon}} = V_{P-\tfrac{\epsilon}{2\beta},\bar{P}- \tfrac{\tilde{\sigma}\epsilon}{2\beta}}^{\sigma, \tilde{\sigma}}\ ,
\end{align}
and the degenerate OPE coefficients $C_{\epsilon}(V)$, $\tilde{C}_{\epsilon}(V)$. 
The existence of OPEs implies that four-point functions can be decomposed into combinations of conformal blocks. 
In particular, an
$s$-channel decomposition is obtained by inserting the OPE of the fields $V_1$ and $V_2$, and reads
\begin{align}
\Big\langle V_1 V_2 V_3 V_4 \Big\rangle = \int\limits_{\mathbb{S}_s} d\mu(V_s)\, D_{s|1234} \mathcal{F}_{\Delta_s}^{(s)}(\Delta_i|z_i)  \mathcal{F}_{\bar\Delta_s}^{(s)}(\bar\Delta_i|\bar z_i)\, ,
\end{align}
where $d\mu(V_s)$ is some measure on the space $\mathbb{S}_s$ of the $s$-channel primary fields. 
This is a combination of four-point conformal blocks $\mathcal{F}_{\Delta_s}^{(s)}(\Delta_i|z_i)$, with the four-point $s$-channel structure constants
\begin{align}\label{eq:dccb}
\begin{tikzpicture}[baseline=(current  bounding  box.center), very thick, scale = .4]
\draw (-1,2) node [left] {$2$} -- (0,0) -- node [above] {$s$} (4,0) -- (5,2) node [right] {$3$};
\draw (-1,-2) node [left] {$1$} -- (0,0);
\draw (4,0) -- (5,-2) node [right] {$4$};
\node at (2.7, -4.4) {$D_{s|1234} = \colorboxed{red}{C_{12}^s}\, \colorboxed{red}{ C_{s34}} =  \displaystyle\frac{C_{12s}C_{s34}}{B_s}\, .$};
\draw[thin, ->, red] (.6,-3.4) to [out=90, in=-70] (.2, -.3);
\draw[thin, ->, red] (3.4,-3.4) to [out=90, in=-110] (3.8, -.3);
\end{tikzpicture} 
\end{align}
Alternatively, we could insert the OPE of the fields $V_1$ and $V_4$, and obtain the $t$-channel decomposition of the same four-point function. Both decompositions should agree, and we obtain the crossing symmetry equation
\begin{align}
\int\limits_{\mathbb{S}_s} d\mu(V_s)\, D_{s|1234}
 \begin{tikzpicture}[baseline=(current  bounding  box.center), very thick, scale = .3]
\draw (-1,2) node [left] {$2$} -- (0,0) -- node [above] {$s$} (4,0) -- (5,2) node [right] {$3$};
\draw (-1,-2) node [left] {$1$} -- (0,0);
\draw (4,0) -- (5,-2) node [right] {$4$}; 
\end{tikzpicture} = \int\limits_{\mathbb{S}_t} d\mu(V_t)\, D_{t|4123}
 \begin{tikzpicture}[baseline=(current  bounding  box.center), very thick, scale = .3]
\draw (-2, -1) node [left] {$1$} -- (0, 0) ; 
\draw (2, -1) node [right] {$4$} -- (0, 0) -- node [right] {$t$} (0, 4) -- (2,5) node [right] {$3$};
\draw (0, 4) -- (-2, 5) node [left] {$2$}; 
\end{tikzpicture}  \label{cross}
\end{align}
Let us now consider a four-point function that involves at least one degenerate field $V_{\langle 2,1 \rangle}$. Since the OPEs of this field have only two terms, its four-point functions involve only two  terms in each channel,
\begin{align}
 \Big\langle V_{\langle 2,1\rangle} V_1 V_2 V_3 \Big\rangle &= 
 \sum_{\epsilon_1 = \pm} d^{(s)}_{\epsilon_1} \mathcal{F}^{(s)}_{\epsilon_1} \bar{\mathcal{F}}^{(s)}_{\sigma_1\epsilon_1}
 = \sum_{\epsilon_3 = \pm} d^{(t)}_{\epsilon_3} \mathcal{F}^{(t)}_{\epsilon_3} \bar{\mathcal{F}}^{(t)}_{\sigma_3\epsilon_3}
 \ ,
 \label{eq:2dec}
\end{align}
where $\mathcal{F}_\epsilon^{(s)},\mathcal{F}^{(t)}_\epsilon$ are degenerate four-point conformal blocks, and the four-point structure constants are of the type 
\begin{align}
\begin{tikzpicture}[baseline=(current  bounding  box.center), very thick, scale = .4]
\draw (-1,2) node [left] {$1$} -- (0,0) -- node [above] {$1^\epsilon$} (4,0) -- (5,2) node [right] {$2$};
\draw[dashed] (-1,-2) node [left] {$\langle 2,1\rangle$} -- (0,0);
\draw (4,0) -- (5,-2) node [right] {$3$};
\node at (2.2, -4) {$d^{(s)}_\epsilon = \colorboxed{red}{C_\epsilon(V_1)}\, \colorboxed{red}{C(V_1^\epsilon,V_2,V_3)} \, .$};
\draw[thin, ->, red] (.6,-3) to [out=90, in=-70] (.2, -.3);
\draw[thin, ->, red] (3.4,-3) to [out=90, in=-110] (3.8, -.3);
\end{tikzpicture} 
\end{align}
Let us introduce the ratio $\rho = \rho(V_1|V_2,V_3) = \frac{d^{(s)}_+}{d^{(s)}_-}$, which is important because we will be able to compute it in Section \ref{sec:rr}. Its expression in terms of structure constants is
\begin{align}
 \rho(V_1|V_2,V_3) = \frac{C_+(V_1) C(V_1^+,V_2,V_3)}{C_-(V_1) C(V_1^-,V_2,V_3)} \ . \label{eq:rcst}
\end{align}
In the particular case of the four-point function $
\left\langle V_{\langle 2,1 \rangle} V_1 V_{\langle 2,1 \rangle} V_1 \right\rangle $, the four-point structure constants are of the type 
\begin{align}
 \begin{tikzpicture}[baseline=(current  bounding  box.center), very thick, scale = .4]
\draw (-1,2) node [left] {$1$} -- (0,0) -- node [above] {$1^\epsilon$} (4,0) -- (5,-2) node [right] {$1$};
\draw[dashed] (-1,-2) node [left] {$\langle 2,1\rangle$} -- (0,0);
\draw[dashed] (4,0) -- (5,2) node [right] {$\langle 2,1\rangle$};
\node at (1, -4.2) {$d^{(s)}_\epsilon = \colorboxed{red}{C_\epsilon(V_1)}\, \colorboxed{red}{B(V_1^\epsilon)}\, \colorboxed{red}{C_\epsilon(V_1)}\ , $};
\draw[thin, ->, red] (-.7,-3.2) to [out=90, in=-70] (.2, -.3);
\draw[thin, ->, red] (4.6,-3.2) to [out=90, in=-110] (3.8, -.3);
\draw[thin, ->, red] (1.9,-3.2) to [out=90, in=-110] (2, -.3);
\end{tikzpicture} 
\end{align}
and the corresponding ratio is
\begin{align}
 \rho(V_1) = \frac{C_+(V_1)^2 B(V_1^+)}{C_-(V_1)^2 B(V_1^-)} \ .
 \label{eq:rvo}
\end{align}
It follows that the dependence of the four-point structure constants $D_{s|1234} = D(V_s)$ \eqref{eq:dccb} on the field $V_s$ is controlled by the shift equations
\begin{align}
\boxed{\frac{D(V_s^+)}{D(V_s^-)} = \frac{\rho(V_s|V_1,V_2)\rho(V_s|V_3,V_4)}{\rho(V_s)}}\, , 
\quad \boxed{\frac{D(V_s^{\tilde{+}})}{D(V_s^{\tilde{-}})} = \frac{\tilde{\rho}(V_s|V_1,V_2)\tilde{\rho}(V_s|V_3,V_4)}{\tilde{\rho}(V_s)}}\, ,
\label{eq:drat}
\end{align}
where the second equation is obtained by replacing $V_{\langle 2,1\rangle}$ with $V_{\langle 1, 2\rangle}$ in our analysis.
Next we will determine the ratios $\rho$ and $\tilde{\rho}$ of degenerate four-point structure constants.

\subsection{Ratios of four-point structure constants}\label{sec:rr}

The $s$- and $t$-channel degenerate four-point conformal blocks $\mathcal{F}_\epsilon^{(s)}$ and $\mathcal{F}^{(t)}_\epsilon$ that appear in the decomposition \eqref{eq:2dec} of $\left\langle V_{\langle 2,1\rangle} V_1 V_2 V_3 \right\rangle$ are two bases of solutions of the same Belavin--Polyakov--Zamolodchikov equation \cite{Ribault:2014hia}. 
They are related by a change of basis of the type 
\begin{align}
\mathcal{F}^{(s)}_{\epsilon_1} = \sum_{\epsilon_3}F_{\epsilon_1,\epsilon_3}\mathcal{F}^{(t)}_{\epsilon_3}\, ,
\end{align} 
whose coefficients are the fusing matrix elements 
\begin{align}
F_{\epsilon_1,\epsilon_3} &= \frac{\Gamma(1+2\beta\epsilon_1 P_1)\Gamma(-2\beta\epsilon_3 P_3)}{\prod_{\pm}\Gamma(\tfrac{1}{2}+\beta\epsilon_1 P_1 \pm \beta P_2-\beta\epsilon_3  P_3))}\ . \label{fusmat}
\end{align} 
The right-moving blocks $\bar{\mathcal{F}}_\epsilon^{(s)}$ and $\bar{\mathcal{F}}^{(t)}_\epsilon$ obey a similar relation, with a fusing matrix $\bar F$ whose elements are obtained from those of $F$ by $P_i\to \bar P_i$.
As a consequence, we obtain four equations relating the structure constants $d^{(s)}_\pm,d^{(t)}_\pm$ of both channels, two for each value of $\epsilon_3 = \pm$:
\begin{align}
 \sum_{\epsilon_1} d^{(s)}_{\epsilon_1} F_{\epsilon_1,\epsilon_3} \bar F_{\sigma_1\epsilon_1,-\sigma_3\epsilon_3} 
 &= 0\, , \label{h21}
 \\
 \sum_{\epsilon_1} d^{(s)}_{\epsilon_1} F_{\epsilon_1,\epsilon_3} \bar F_{\sigma_1\epsilon_1,\sigma_3\epsilon_3}
 &= d^{(t)}_{\epsilon_3}\label{ih21}\, .
\end{align}
The four-point function vanishes unless equations \eqref{h21} admit a non-zero solution for $d^{(s)}_{\epsilon}$, which happens only if
\begin{align}
 \frac{F_{++}F_{--}}{F_{+-}F_{-+}} = \left(\frac{\bar{F}_{++}\bar{F}_{--}}{\bar{F}_{+-}\bar{F}_{-+}} \right)^{\sigma_1\sigma_3}\, . \label{cons21}
\end{align}
Explicitly, this condition reads
\begin{align}
 \prod_{\pm} \frac{\cos \pi \beta (P_1\pm P_2-P_3)}{\cos \pi \beta (P_1\pm P_2 + P_3)} = \prod_\pm \frac{\cos \pi \beta (\sigma_1\bar P_1\pm \bar P_2 - \sigma_3 \bar P_3)}{\cos \pi \beta (\sigma_1\bar P_1\pm \bar P_2 +\sigma_3\bar P_3)} \, . 
\end{align}
Assuming that our three fields are diagonal or non-diagonal as discussed in Section \ref{sec:dnd}, 
the numbers $s_i =  \beta (\sigma_i \bar{P}_i-P_i)$ must be half-integer. (For a non-diagonal field, $s_i$ coincides with the half-integer index of a same name. For a diagonal field, we have $s_i=0$.) Then our condition reduces to 
\begin{align}
 \left(1-(-1)^{2\sum_{i=1}^3 s_i}\right)\sin(2\pi\beta P_1)\cos(2\pi \beta P_2)\sin(2\pi \beta P_3) = 0\ .
\end{align}
Assuming we are in the generic situation where the trigonometric factors do not vanish, this implies
\begin{align}
 \boxed{\sum_{i=1}^3 s_i \in \mathbb{Z}}\, .
 \label{eq:sums}
\end{align}
Now, since fusion with $V_{\langle 2,1\rangle}$ \eqref{dfus} leaves
the number $s$ unchanged, this condition holds not only for our four-point function $\left\langle V_{\langle 2,1\rangle} V_1 V_2 V_3 \right\rangle $ , but also for the three-point functions that result from the fusion $V_{\langle 2,1\rangle} \times V_1$. 
It follows that any nonzero three-point function must obey this condition. 
The same analysis with the  correlation function $\left\langle V_{\langle 1,2 \rangle}V_1V_2V_3 \right\rangle $, and with the convention $r=0$ for a diagonal field, leads to the analogous condition
\begin{align}
\boxed{\sum_{i=1}^3 r_i \in \mathbb{Z}}\, .\label{eq:sumr}
\end{align}
To illustrate the  implications of these conditions, let us consider two examples:
\begin{itemize}
\item For any three-point function of the type $\left<V^DV^DV^N\right>$, the non-diagonal field must have integer indices $r,s\in\mathbb{Z}$. 
\item Any three-point function with an odd number of fermionic fields vanishes, because fermionic fields obey $r_i+s_i\in\mathbb{Z}+\frac12$ while $\sum_{i=1}^3 (r_i+s_i)\in\mathbb{Z}$. 
So our conditions imply the single-valuedness condition \eqref{sumsp} for three-point functions.
\end{itemize}
Returning to the four-point function $\left\langle V_{\langle 2,1\rangle} V_1 V_2 V_3 \right\rangle $  we see that, if conditions \eqref{cons21} are obeyed, the ratios of the four-point structure constants are given in terms of fusing matrix elements,
\begin{align}
\rho = \frac{d^{(s)}_{+}}{d^{(s)}_{-}} = -\frac{F_{-,\epsilon_3} \bar F_{-\sigma_1,-\sigma_3\epsilon_3}  }{ F_{+,\epsilon_3} \bar F_{\sigma_1,-\sigma_3\epsilon_3}}\, ,\quad  \forall \epsilon_3\in\{+,-\}\ .  \label{eq:rfus}
\end{align}
Inserting explicit expressions \eqref{fusmat} for the fusing matrix elements, we obtain
\begin{multline}
 \rho(V_1|V_2,V_3) = -  \frac{\Gamma(-2\beta P_1)}{\Gamma(2\beta P_1)} \frac{\Gamma(-2\beta \sigma_1\bar P_1)}{\Gamma(2\beta \sigma_1\bar P_1)} 
 \\
 \times
 \prod_\pm \frac{\Gamma(\frac12+\beta P_1\pm \beta P_2 +\beta \epsilon_3P_3)}{\Gamma(\frac12 -\beta P_1 \pm \beta P_2 +\beta \epsilon_3 P_3)}
 \prod_\pm \frac{\Gamma(\frac12+\beta \sigma_1\bar P_1\pm \beta \bar P_2 -\beta \epsilon_3\sigma_3\bar P_3)}{\Gamma(\frac12 -\beta \sigma_1\bar P_1\pm \beta \bar P_2-\beta \epsilon_3\sigma_3\bar P_3)}\ ,
\end{multline}
where we have restored the explicit dependence of $\rho=\rho(V_1|V_2,V_3)$ on the fields.
Remembering that $s_i =  \beta(\sigma_i \bar{P}_i-P_i)\in\frac12 \mathbb{Z}$ obey eq. \eqref{eq:sums},
this can be rewritten in manifestly $\epsilon_3,\sigma_3$-independent way:
\begin{align}
 \boxed{\rho(V_1|V_2,V_3) = -(-1)^{2s_2} \frac{\Gamma(-2\beta P_1)}{\Gamma(2\beta P_1)} \frac{\Gamma(-2\beta \sigma_1\bar P_1)}{\Gamma(2\beta \sigma_1\bar P_1)}  \frac{\prod_{\pm, \pm} \Gamma(\frac12 +\beta P_1 \pm \beta P_2\pm \beta P_3)}{\prod_{\pm, \pm} \Gamma(\frac12 -\beta \sigma_1\bar P_1 \pm \beta \bar P_2 \pm \beta \bar P_3)}}\, .
 \label{eq:r}
\end{align}
Let us check that this behaves as expected when we permute the fields $V_2$ and $V_3$. Using eq. \eqref{eq:sums}, we find 
\begin{align}
 \rho(V_1|V_2,V_3)=(-1)^{2s_1}\rho(V_1|V_3,V_2)\ .
\end{align}
This is actually what we expect from the expression of $\rho(V_1|V_2,V_3)$ in terms of structure constants \eqref{eq:rcst}, given their behaviour \eqref{permC} under permutations, and the relation $(-1)^{2s_1} = (-1)^{S(V_1^+)-S(V_1^-)}$.

The analogous expression for the ratio $\tilde{\rho}=\frac{\tilde{d}^{(s)}_{+}}{\tilde{d}^{(s)}_{-}}$ of $s$-channel structure constants of the four-point function $\left\langle V_{\langle 1,2\rangle} V_1 V_2 V_3 \right\rangle $  is obtained by the substitutions $s\to r,\ \sigma\to \tilde{\sigma},\ \beta\to -\beta^{-1}$:
\begin{multline}
 \tilde{\rho}(V_1|V_2,V_3) = -(-1)^{2r_2} \frac{\Gamma(2\beta^{-1}P_1)}{\Gamma(-2\beta^{-1}P_1)} \frac{\Gamma(2\beta^{-1}\tilde{\sigma}_1\bar P_1)}{\Gamma(-2\beta^{-1}\tilde{\sigma}_1\bar P_1)} 
 \\
 \times
 \frac{ \prod_{\pm, \pm} \Gamma(\frac12 -\beta^{-1}P_1 \pm \beta^{-1}P_2\pm \beta^{-1}P_3)}{ \prod_{\pm, \pm} \Gamma(\frac12 +\beta^{-1}\tilde{\sigma}_1\bar P_1 \pm \beta^{-1}\bar P_2 \pm \beta^{-1}\bar P_3)}\ .
 \label{eq:rt}
\end{multline}
The particular case of a four-point function $
\left\langle V_{\langle 2,1 \rangle} V V_{\langle 2,1 \rangle} V \right\rangle $ with two degenerate fields  corresponds to 
\begin{align}
(P_1,\bar{P}_1) &= (P_3,\bar{P}_3)  = (P, \bar{P})\, ,\\
P_2 = \bar{P}_2 &= \beta-\tfrac{1}{2\beta}\, , 
\end{align} 
and we find 
\begin{align}
 \boxed{\rho(V) =  -\frac{\Gamma(-2\beta P)\Gamma(-2\beta \sigma \bar{P})}{\Gamma(2\beta P)\Gamma(2\beta \sigma \bar{P})}\frac{\Gamma(\beta^2 +2\beta P)\Gamma(1-\beta^2 +2\beta P)}{\Gamma(\beta^2 -2\beta\sigma\bar{P})\Gamma(1-\beta^2-2\beta\sigma\bar{P})}}\, . \label{OPEcg}
\end{align}
Similarly, the ratio of structure constants of the four-point function $\left\langle V_{\langle 1,2 \rangle} V V_{\langle 1,2 \rangle} V \right\rangle $ is
\begin{align}
 \tilde{\rho}(V) =  -\frac{\Gamma(2\beta^{-1} P)\Gamma(2\beta^{-1} \tilde{\sigma} \bar{P})}{\Gamma(-2\beta^{-1} P)\Gamma(-2\beta^{-1} \tilde{\sigma} \bar{P})}\frac{\Gamma(\beta^{-2} -2\beta^{-1} P)\Gamma(1-\beta^{-2} -2\beta^{-1} P)}{\Gamma(\beta^{-2} +2\beta^{-1}\tilde{\sigma}\bar{P})\Gamma(1-\beta^{-2}+2\beta^{-1}\tilde{\sigma}\bar{P})}\, . \label{OPEcgt}
\end{align}
These ratios slightly simplify in the case of diagonal fields i.e. if $P=\sigma \bar P = \tilde{\sigma}\bar P$,
\begin{align}
 \rho(V^D_P) = -\frac{\Gamma^2(-2\beta P)}{\Gamma^2(2\beta P)} \frac{\gamma(\beta^2+2\beta P)}{\gamma(\beta^2-2\beta P)} \, , \quad
 \tilde{\rho}(V^D_P) = -\frac{\Gamma^2(2\beta^{-1} P)}{\Gamma^2(-2\beta^{-1} P)} \frac{\gamma(\beta^{-2}-2\beta^{-1} P)}{\gamma(\beta^{-2}+2\beta^{-1} P)}\, , \label{OPEcd}
\end{align}
where we introduced $\gamma(x)=\frac{\Gamma(x)}{\Gamma(1-x)}$.

The determination of the ratios $\rho,\tilde{\rho}$ lead to shift equations \eqref{eq:drat} for the four-point structure constants that appear in generic four-point functions $\left< V_1V_2V_3V_4 \right>$.  For some spectrums, these shift equations are enough for determining the structure constants up to a $V_s$-independent factor, and therefore enough for checking crossing symmetry. We will see examples of this in Section \ref{sec:csfp}. For the moment, let us determine three-point structure constants.

\subsection{Three-point structure constants}\label{ssec:3sc}

Before studying how our conformal bootstrap equations constrain three-point structure constants, let us warn that in principle such structure constants cannot be fully determined.
This is because our fields, and therefore also their correlation functions, are only defined up to a $z$-independent renormalization,
\begin{align}
 V_i(z) \to \lambda_i V_i(z)\, . \label{eq:renorm}
\end{align}
We could fix this ambiguity by imposing additional constraints on correlation functions, as is done in minimal models by imposing $B(V)=1$, or in Liouville theory by imposing $C_+(V)=\tilde{C}_+(V)=1$ \cite{Ribault:2014hia}. However, in the case of non-diagonal fields, imposing either constraint would lead to correlation functions having factors that involve square roots of Gamma functions. Such factors would be complicated, and non-analytic in $\beta$.
Rather, we will introduce a reference normalization where the three-point function has a particularly simple expression $C'_{123}$, and write the three-point structure constant in an arbitrary normalization as
\begin{align}
 C_{123} = \left(\prod_{i=1}^3 Y_i\right)C'_{123}\, , \label{3pnorm}
\end{align}
Then $C'_{123}$ is normalization-independent,  and  the normalization factor $Y_i$  behaves as $Y_i\to \lambda_i Y_i$ under \eqref{eq:renorm}. 
Since the two-point structure constant \eqref{2pt} behaves as $B_i \to \lambda_i^2 B_i$, the normalization-independent quantities that we can hope to determine are
\begin{align}
 C'_{123} \qquad \text{and} \qquad Y_i^2B_i^{-1}\ .
\end{align}
Rewriting the four-point $s$-channel structure constant as 
\begin{align}
 D_{s|1234} = \left(\prod_{i=1}^4 Y_i\right) C'_{12s} C'_{s34} \frac{Y^2_s}{B_s}\, , \label{eq:4pcp}
\end{align}
shows that the crossing symmetry equation \eqref{cross} only involves normalization-independent quantities.

The equations that constrain the structure constants are written in terms of the ratios $\rho$ \eqref{eq:rcst} and $\tilde{\rho}$. If we define the normalization-independent ratio
\begin{align}
\rho'(V_1|V_2,V_3)= \frac{C'(V_1^{+},V_2,V_3)}{C'(V_1^{-},V_2,V_3)}\, , \label{eq:rp}
\end{align}
then the combination
\begin{align}
\frac{\rho(V_1|V_2,V_3)}{\rho'(V_1|V_2,V_3)} = \frac{C_{+}(V_1)Y(V_1^{+})}{C_{-}(V_1)Y(V_1^{-})} \, , 
\label{eq:rrp}
\end{align}
should only depend on the field $V_1$. 
So let us look for a function $C'(V_1,V_2,V_3)$ such that the corresponding combinations $\frac{\rho}{\rho'}$ and $\frac{\tilde{\rho}}{\tilde{\rho}'}$ only depend on $V_1$. 
The problem simplifies significantly if at least one of the fields is diagonal, say $V_1=V^D_{P_1}$, and we will restrict to this case. We propose the ansatz
\begin{align}
\boxed{C'\left(V^{D}_{P_1}, V_2, V_3\right) = \frac{ f_{2,3}(P_1) }{ \prod\limits_{\pm, \pm} 
\Gamma_\beta(\frac{\beta}{2}+\frac{1}{2\beta} +P_1\pm P_2 \pm P_3)
\prod\limits_{\pm, \pm} \Gamma_\beta(\frac{\beta}{2}+\frac{1}{2\beta}-P_1\pm \bar{P}_2 \pm \bar{P}_3)}}\, , \label{cdnn}
\end{align}
where $\Gamma_\beta(x)$ is a double Gamma function with periods $\beta$ and $\beta^{-1}$, which is invariant under $\beta\to \beta^{-1}$ and obeys 
\begin{align}
 \Gamma_\beta(x+\beta) = \sqrt{2\pi}\frac{\beta^{\beta x-\frac12}}{\Gamma(\beta x)} \Gamma_\beta(x)\ .
\end{align}
The factor $f_{2,3}(P_1)\in\{-1,+1\}$ is included in order to account for the sign factors appearing in eqs. \eqref{eq:r} and \eqref{eq:rt}. It is determined by the equations
\begin{align}
\frac{ f_{2,3}(P_1 +\frac{\beta}{2})}{f_{2,3}(P_1 -\frac{\beta}{2})} = (-1)^{2s_2} \quad , \quad \frac{f_{2,3}(P_1 +\frac{1}{2\beta})}{ f_{2,3}(P_1 -\frac{1}{2\beta})} =(-1)^{2r_2}\, . \label{eq:f23}
\end{align}
If $r_2,s_2\in \mathbb{Z}$, in particular if either $V_2$ or $V_3$ is diagonal, we have $f_{2,3}(P_1) = 1$. 
With our ansatz, the ratios are
\begin{align}
\rho'(V^D_1|V_2,V_3) &= (-1)^{2s_2} \beta^{-8\beta P_1}  \frac{\prod_{\pm, \pm} \Gamma\left(\frac{1}{2}+\beta P_1\pm \beta P_2 \pm \beta P_3\right)}{\prod_{\pm, \pm}\Gamma\left(\frac{1}{2}-\beta P_1\pm \beta \bar P_2 \pm \beta \bar P_3\right)}\, , \\
\tilde{\rho}'(V^D_1|V_2,V_3) &= (-1)^{2r_2} \beta^{-\frac{8}{\beta}P_1}  \frac{\prod_{\pm, \pm}\Gamma\left(\frac{1}{2}-\beta^{-1} P_1\pm \beta^{-1}\bar{P}_2 \pm \beta^{-1}\bar{P}_3\right)}{\prod_{\pm, \pm}\Gamma\left(\frac{1}{2}+ \beta^{-1}P_1\pm \beta^{-1}P_2 \pm \beta^{-1}P_3\right)}\, ,
\end{align}
which indeed match the expressions \eqref{eq:r} and \eqref{eq:rt} for $\rho(V_1|V_2,V_3)$ and $\tilde{\rho}(V_1|V_2,V_3)$, up to factors that depend on $V_1$ only. 

Notice that the denominator of $C'(V^D_1, V_2, V_3)$ \eqref{cdnn} is invariant under $P_1 \to -P_1$. 
Neglecting the numerator, we indeed have 
\begin{align}
\frac{C'(V^D_{P_1}, V_2, V_3)}{C'(V^D_{-P_1}, V_2, V_3)} \propto
 \frac{\prod\limits_{\pm, \pm} S_\beta(\frac{\beta}{2}+\frac{1}{2\beta}+P_1\pm \bar{P}_2 \pm \bar{P}_3)}{\prod\limits_{\pm, \pm} S_\beta(\frac{\beta}{2}+\frac{1}{2\beta} +P_1\pm P_2 \pm P_3)} \ ,
\end{align}
with $S_\beta(x) = \frac{\Gamma_\beta(x)}{\Gamma_\beta(\beta+\frac{1}{\beta}-x)}$.
Using the shift equation $\frac{S_\beta(x+\beta)}{S_\beta(x)} = 2 \sin(\pi\beta x)$, we find the identity 
\begin{align}
\prod_\pm \frac{S_\beta(x\pm P_{(r,-s)})}{S_\beta(x \pm P_{(r,s)})}=(-1)^{rs}, \quad \forall r,s \in \mathbb{Z} \, . \label{eq:IdS}
\end{align}
We can use this identity 
thanks to the non-triviality conditions \eqref{eq:sums} and \eqref{eq:sumr}, and we find 
\begin{align}
 \frac{C'(V^D_{P_1}, V_2, V_3)}{C'(V^D_{-P_1}, V_2, V_3)} \propto (-1)^{2(r_2s_2 + r_3s_3)} = 1\ ,
\end{align}
where we also used the single-valuedness of correlation functions eq. \eqref{sumsp}.

Using the combination $\frac{\rho}{\rho'}$ \eqref{eq:rrp},
together with the ratio $\rho(V_1)$ \eqref{eq:rvo}, 
we can write a shift equation for the normalization-independent quantity $Y^2B^{-1}$,
\begin{align}
\frac{Y^{2}B^{-1}(V^{+}_{1})}{Y^{2}B^{-1}(V^{-}_{1})} = \frac{\rho^2(V_1|V_2,V_3)}{\rho'^2(V_1|V_2,V_3)\rho(V_1)}\, . \label{eq:YBr}
\end{align}
In the case of a diagonal field, this equation is explicitly
\begin{align}
\frac{(Y^2B^{-1})\left(V^{D,+}_{1}\right)}{(Y^2B^{-1})\left(V^{D,-}_{1}\right)} 
= 
-\beta^{16\beta P_1}\frac{\Gamma^2(-2\beta P_1)}{\Gamma^2(2\beta P_1)} \frac{\gamma(\beta^2-2\beta P_1)}{\gamma(\beta^2+2\beta P_1)}\, ,
\end{align}
and the dual shift equation is 
\begin{align}
\frac{(Y^2B^{-1})\left(V^{D,\tilde{+}}_{1}\right)}{(Y^2B^{-1})\left(V^{D,\tilde{-}}_{1}\right)} 
= 
-\beta^{\frac{16}{\beta }P_1}\frac{\Gamma^2(2\beta^{-1}P_1)}{\Gamma^2(-2\beta^{-1}P_1)} \frac{\gamma(\beta^{-2}+2\beta^{-1}P_1)}{\gamma(\beta^{-2}-2\beta^{-1}P_1)}\, . \label{dseY2B}
\end{align}
A solution can be written in terms of the function $\Upsilon_\beta(x) = \frac{1}{\Gamma_\beta(x)\Gamma_\beta(\beta+\beta^{-1}-x)}$ as
\begin{align}
(Y^2B^{-1})(V^D_P) = \frac{1}{\prod_\pm \Upsilon_\beta(\beta\pm 2P)} \, .\label{eq:YBUps}
\end{align}
Let us now determine the factor $Y^2B^{-1}$ for a field $V_2$ that is not necessarily diagonal. We set $\sigma_2 = -\tilde{\sigma}_2 =1$, but the result does not depend on these choices. We compute the ratios $\rho(V_2|V_3, V^D_1)$ and $\tilde{\rho}(V_2|V_3, V^D_1)$ by renaming the momentums in \eqref{eq:r} and \eqref{eq:rt}. Making use of the permutation properties of the three-point structure constant \eqref{permC} we rewrite
\begin{align}
\rho'(V_2|V_3, V^D_1)= \frac{C'(V^D_1, V^{+}_2, V_3)}{C'(V^D_1, V^{-}_2, V_3)}\, , \quad \tilde{\rho}'(V_2|V_3, V^D_1)= \frac{C'(V^D_1, V^{\tilde{+}}_2, V_3)}{C'(V^D_1, V^{\tilde{-}}_2, V_3)}\ .
\end{align}
Using the ansatz \eqref{cdnn}, we compute these ratios. Inserting them into eq. \eqref{eq:YBr}, we obtain
\begin{multline}
\frac{(Y^2B^{-1})\left(V^{+}_{2} \right) }{(Y^2B^{-1})\left(V^{-}_{2} \right)}= -\beta^{8\beta(P_2+\bar{P}_2)} \frac{\Gamma(-2\beta P_2)\Gamma(-2\beta\bar{P}_2)}{\Gamma(2\beta P_2)\Gamma(2\beta\bar{P}_2)}
\\ 
\times
\frac{\Gamma(\beta^2-2\beta\bar{P}_2)\Gamma(1-\beta^2-2\beta\bar{P}_2)}{\Gamma(\beta^2+2\beta P_2)\Gamma(1-\beta^2+2\beta P_2)}\, ,
\end{multline}
\begin{multline}
\frac{(Y^2B^{-1})\left(V^{\tilde{+}}_{2} \right) }{(Y^2B^{-1})\left(V^{\tilde{-}}_{2} \right)}
= 
-\beta^{8\beta^{-1}(P_2-\bar{P}_2)} 
\frac{\Gamma(2\beta^{-1}P_2)\Gamma(-2\beta^{-1}\bar{P}_2)}{\Gamma(-2\beta^{-1}P_2)\Gamma(2\beta^{-1}\bar{P}_2)} 
\\ 
\times
\frac{\Gamma(\beta^{-2}-2\beta^{-1}\bar{P}_2)\Gamma(1-\beta^{-2}-2\beta^{-1}\bar{P}_2)}{\Gamma(\beta^{-2}-2\beta^{-1}P_2)\Gamma(1-\beta^{-2}-2\beta^{-1}P_2)} \, .
\end{multline}
A solution of these equations is 
\begin{align}
\boxed{ (Y^{2}B^{-1})\left( V_{P,\bar P}^{\sigma,\tilde{\sigma}} \right) =(-1)^{P^2-\bar P^2} \prod_{\pm} \Gamma_\beta(\beta\pm 2P)\Gamma_\beta(\beta^{-1}\pm 2\bar{P})}\, , \label{Y2Bnd}
\end{align}
which reduces to the diagonal case expression \eqref{eq:YBUps} if $P = \pm \bar{P}$. 

So we have found solutions of the shift equations for the normalization-independent quantities $C'$ and $Y^2B^{-1}$, with the exception of three-point structure constants that involve three non-diagonal fields. 
The formulas are slightly more complicated in that case, and we leave them for future work. Notice however that the shift equations \eqref{eq:drat} are enough for practical purposes, as we will demonstrate in an example in Section \ref{sec:atm}.

To conclude, let us discuss whether our solution $C'$ of the shift equations is unique. The dependences on diagonal and non-diagonal fields deserve separate discussions:
\begin{itemize}
 \item Just as in the case of Liouville theory \cite{rib16}, shift equations uniquely determine the dependence of three-point structure constants on momentums of a \textbf{diagonal field}  $V_P^D$ provided $\beta\in\mathbb{R}$ or $\beta \in i\mathbb{R}$. Just like the function $\Gamma_\beta$, our solution is well-defined and analytic for $\beta\notin i\mathbb{R}$, and is therefore the unique analytic continuation of the $\beta\in\mathbb{R}$ solution to the domain $\beta \notin i\mathbb{R}$ i.e. $c\notin (25,\infty)$. 
 \item When it comes to a \textbf{non-diagonal field} $V^N_{(r,s)}$, the shift equations determine how three-point structure constants change when each one of the indices $r,s$ is shifted by two units. Given the allowed values \eqref{eq:rsval} of these indices, the space of solutions of the shift equations is finite-dimensional. In specific models however, the indices do not take all their allowed values. In the case of the spectrum \eqref{eq:dlim}, shift equations plus invariance under the reflection $(r,s)\to (-r,-s)$ have a unique solution. In the case of the Ashkin--Teller model (Section \ref{sec:atm}), the space of solutions is two-dimensional.
\end{itemize}

\subsection{Relation with Liouville theory}\label{sec:liou}

Let us investigate whether our structure constants obey a relation of the type \eqref{eq:csqrt} with structure constants of Liouville theory.
Let us start with the case of a three-point structure constant of three diagonal fields. 
In this case, the normalization-independent factor \eqref{cdnn} reduces to
\begin{align}
 C'(V^D_1,V^D_2,V^D_3) = \prod_{\pm, \pm}\Upsilon_\beta\left(\tfrac{\beta}{2}+\tfrac{1}{2\beta}+P_1\pm P_2\pm P_3\right)\ .
 \label{eq:cpl}
\end{align}
For $c\leq 1$ (i.e. $\beta\in\mathbb{R}$) this coincides with the analogous quantity in Liouville theory, $C_{\text{L}}(P_1, P_2, P_3)$, as reviewed for example in \cite{Ribault:2014hia}. 
For other values of $c$ though, we have to use a different solution of the shift equations in order to recover Liouville theory results. 
Similarly, our solution for $(Y^2B^{-1})(V_P^D)$ \eqref{eq:YBUps} agrees with Liouville theory only if $c\leq 1$, and has to be replaced with the solution $(Y^2B^{-1})_\text{L}(P) \underset{c\notin (-\infty,1)}{=} \prod_\pm \Upsilon_{i\beta}(\pm 2iP)$ otherwise. 
We will not elaborate on this subtlety, because it affects only diagonal fields with their continuous values of the momentum $P$. 
Liouville theory structure constants that involve at least one discrete momentum of the type $P_{(r,s)}$ are determined by shift equations, modulo finitely many initial conditions. 
This implies in particular that they are analytic functions of $\beta$, with no singularity at $\beta\in\mathbb{R}$. 

So, if a least one momentum is discrete, we can use the $\beta\in\mathbb{R}$ Liouville theory formula \eqref{eq:cpl} for complex values of $\beta$ as well. Then it is straightforward to check that our normalization-independent factor \eqref{cdnn} obeys
\begin{align}
C'^2(V^D_1|V_2,V_3) = C'_{\text{L}}(P_1,P_2,P_3) C'_{\text{L}}(P_1,\bar{P}_2,\bar{P}_3)\, . \label{eq:cn2cd}
\end{align}
Let us check the analogous relation for the quantity $Y^2B^{-1}$ \eqref{Y2Bnd}. We compute
\begin{align}
\frac{(Y^2B^{-1})\left(V^N_{(r,s)}\right)}{(Y^2B^{-1})\left(V^N_{(r,-s)}\right)} =\prod_\pm \frac{S_\beta(\beta\pm 2P_{(r,-s)})}{S_\beta(\beta \pm 2P_{(r,s)})}
=(-1)^{4rs}=1 \, , \label{eq:Yrat}
\end{align}
where we used the identity \eqref{eq:IdS}.
This implies 
\begin{align}
 \left[(Y^2B^{-1})\left(V^N_{(r,s)}\right)\right]^2 = (Y^2B^{-1})\left(V^N_{(r,s)}\right)(Y^2B^{-1})\left(V^N_{(r,-s)}\right)\ .
\end{align}
Using $\bar P_{(r,s)} = P_{(r,-s)}$ we obtain
\begin{align}
 \left[(Y^2B^{-1})\left(V^N_{(r,s)}\right)\right]^2 
=  (Y^2B^{-1})\left(V^D_{P_{(r,s)}}\right) (Y^2B^{-1})\left(V^D_{\bar P_{(r,s)}}\right)\, . \label{eq:Y2Yd}
\end{align}
Together with eq. \eqref{eq:cn2cd}, this shows that the geometric mean relation \eqref{eq:csqrt} holds at the level of normalization-independent quantities. We have actually written the square of this relation, in order to avoid having sign ambiguities. These sign ambiguities make the geometric mean relation more suggestive than practically useful. Still,  squares of three-point structure constants do appear in four-point structure constants of the type $D_{s|1212}$, and we find
\begin{align}
D_{s|1212} = Y^2(V_1)Y^2(V_2) (Y^2B^{-1})(V_s) C'_{\text{L}}(P_1,P_2,P_s) C'_{\text{L}}(\bar{P}_1,\bar{P}_2,\bar{P}_s)\, . \label{eq:d12}
\end{align}
This shows that the crossing symmetry equations for four-point functions of the type $\left<V_1V_2V_1V_2\right>$ can be written in terms of Liouville three-point structure constants.

Our structure constants, and the shift equations that they solve, are ultimately derived from the fusing matrix. 
Therefore, the relation with Liouville theory should be expressible in terms of this matrix. 
Let us indeed write the square of the ratio \eqref{eq:rfus} of fusing matrix elements, as the product of the two equivalent expressions for this ratio:
\begin{align}
 \rho^2 = \prod_{\epsilon_3 = \pm} \frac{ F_{+,\epsilon_3} \bar F_{\sigma_1,-\sigma_3\epsilon_3}}{F_{-,\epsilon_3} \bar F_{-\sigma_1,-\sigma_3\epsilon_3}  } = \frac{F_{++}F_{+-}}{F_{-+}F_{--}} \left(\frac{\bar{F}_{++}\bar{F}_{+-}}{\bar{F}_{-+}\bar{F}_{--}} \right)^{\sigma_1} \ .
\end{align}
This relation can be rewritten as 
\begin{align}
 \rho^2(V_1|V_2, V_3) = \rho_{\text{L}}(P_1|P_2,P_3) \rho^{\sigma_1}_{\text{L}}(\bar{P}_1|\bar{P}_2,\bar{P}_3)\, ,
 \label{rhrh}
\end{align}
where $\rho_\text{L}=\frac{F_{++}F_{+-}}{F_{-+}F_{--}}  $ is the expression for $\rho$ when all three fields are diagonal.
This relation implies the geometric mean relation for shift equations and therefore for structure constants. 
See Appendix \ref{app} for a discussion of how general the geometric mean relation might be.

\section{Crossing-symmetric four-point functions}\label{sec:csfp}

In order to compute a four-point function, we need not only structure constants, but also a spectrum. 
But the spectrum of an OPE of two non-degenerate fields is a priori not easy to determine. 
In order to find plausible guesses for the spectrums of some OPEs, we will start with known OPEs in minimal models, and send the central charge to non-rational values.

\subsection{Non-rational limit of minimal models}\label{ssec:mmlim}

Let us first illustrate our approach in the case of diagonal models. 
For any coprime integers $p,q\geq 2$, there exists a diagonal (A-series) minimal model with the parameter $\beta^2 = \frac{p}{q}$.
Its spectrum is built from degenerate representations $\mathcal{R}_{\langle r,s\rangle}$ of the Virasoro algebra,
\begin{align}
 \mathcal{S}^{\text{A-series}}_{p,q} = \frac12 \bigoplus_{r=1}^{q-1}\bigoplus_{s=1}^{p-1} \left| \mathcal{R}_{\langle r,s\rangle} \right|^2\ ,
\end{align}
where by $|\mathcal{R}|^2=\mathcal{R}\otimes \bar{\mathcal{R}}$ we mean a representation $\mathcal{R}$ of the left-moving Virasoro algebra, tensored with the same representation of the right-moving Virasoro algebra.
Let us consider the limit of this spectrum as $p,q\to \infty$ such that $\frac{p}{q}\to \beta_0^2$, for a given real value of $\beta_0$. Assuming $\beta_0^2\notin\mathbb{Q}$, the momentums $P_{(r,s)}$ \eqref{eq:prs} of the states become dense in the real line. Moreover, for a generic momentum $P_0\in\mathbb{R}$, we have $P_{(r,s)}\to P_0 \implies r,s\to\infty$. It follows that the levels of the null vectors of $\mathcal{R}_{\langle r,s\rangle}$ go to infinity, and that $\underset{P_{(r,s)}\to P_0}{\lim} \mathcal{R}_{\langle r,s\rangle} = \mathcal{V}_{P_0}$, where $\mathcal{V}_{P_0}$ is the Verma module of momentum $P_0$.
To summarize, the minimal models' spectrums tend to a diagonal, continuous spectrum made of Verma modules,
\begin{align}
 \boxed{\underset{\frac{p}{q}\to \beta_0^2}{\lim} \mathcal{S}^{\text{A-series}}_{p,q} = \int_{\mathbb{R}_+} dP\, \left|\mathcal{V}_P\right|^2}\ .
\end{align}
We thus recover the spectrum of Liouville theory. Since the three-point structure constants of Liouville theory and minimal models are analytic in $\beta,P$ and obey the same shift equations, we conjecture that the correlation functions of Liouville theory with $\beta\in\mathbb{R}$ i.e. $c\leq 1$ are limits of correlation functions of diagonal minimal models. 
Notice however that the spectrum of Liouville theory with $c\leq 1$ was found much later than its structure constants \cite{rs15}. Taking limits of minimal models is a shortcut that would have led to the correct spectrum, and that we will now use in the case of non-diagonal models.

For any coprime integers $p,q$ such that $q\geq 6$ is even and $p\geq 3$ is odd, there exists a non-diagonal (D-series) minimal model with the parameter $\beta^2 = \frac{p}{q}$ \cite{fms97}. 
The spectrum can be split into a diagonal and a non-diagonal sector, 
\begin{align}
 \mathcal{S}_{p,q}^{\text{D-series}} \ =\  \frac12 \bigoplus_{r\overset{2}{=}1}^{q-1} \bigoplus_{s=1}^{p-1} \left|\mathcal{R}_{\langle r,s\rangle}\right|^2 \oplus \frac12 \ \bigoplus_{\substack{1\leq r\leq q-1 \\ r\equiv \frac{q}{2}\bmod 2}} \  \bigoplus_{s=1}^{p-1} \mathcal{R}_{\langle r,s\rangle} \otimes \bar{\mathcal{R}}_{\langle q-r,s\rangle}\ .
\end{align}
The notation $\overset{2}{=}$ means that $r$ increases in steps of $2$, implying that the diagonal representations have $r$ odd, while the non-diagonal representations have $r \equiv \frac{q}{2} \bmod 2$. 
If $q \equiv 2 \bmod 4$ the representation $\big|\mathcal{R}_{\langle \frac{q}{2},s\rangle}\big|^2$ appears with multiplicity $2$, with one copy in the diagonal sector and the other copy in the non-diagonal sector. If $q \equiv 0 \bmod 4$ every representation has multiplicity $1$.
 
The fusion rules of the D-series minimal models have the remarkable property that diagonality is conserved \cite{run99},
\begin{align}
 D\times D = D \quad , \quad D \times N = N \quad , \quad N \times N = D \ .
 \label{eq:dnf}
\end{align}
This property, plus the fusion products of degenerate representations, are enough for determining the spectrums of the OPEs in D-series minimal models.

In the non-diagonal sector of the spectrum $\mathcal{S}_{p, q}^{\text{D-series}}$, spins take the values
\begin{align}
S = \Delta_{( q-r, s)} - \Delta_{(r,s)} = \left(r-\frac{q}{2}\right)\left(s-\frac{p}{2}\right)\ .
\end{align}
These spins are integer, and must therefore remain constant when we take a limit $p,q\to \infty$ such that $\frac{p}{q}\to \beta_0^2$.
This suggests that we take both factors $r-\frac{q}{2}$ and $s-\frac{p}{2}$ to be constant, and assume that $r,s$ take $p,q$-dependent values, 
\begin{align}
 \left\{\begin{array}{l} r = \frac{q}{2} + r_0\ , \\ s = \frac{p}{2} + s_0\ , \end{array}\right. \quad \text{with} \quad  \left\{\begin{array}{l} r_0\in 2\mathbb{Z} \ , \\ s_0 \in \mathbb{Z} + \frac12\ . \end{array}\right.
\end{align}
Then $\Delta_{( r, s)} = \Delta_{( r_0, s_0)}$ and $\Delta_{( q-r, s)} = \Delta_{( r_0, -s_0)}$, so not only the spins but also the left and right dimensions of our states remain constant in our limit. 
On the other hand, the diagonal sector behaves just like the spectrum of a diagonal minimal model in this limit. 
To summarize,
\begin{align}
 \boxed{\underset{\frac{p}{q}\to \beta_0^2}{\lim} \mathcal{S}^{\text{D-series}}_{p,q} = \frac12 \int_{\mathbb{R}} dP\, \left|\mathcal{V}_P\right|^2 \oplus \mathcal{S}_{2\mathbb{Z},\mathbb{Z}+\frac12}}\ ,
 \label{eq:dlim}
\end{align}
where we use the notation $\mathcal{S}_{X,Y} = \bigoplus_{r\in X}\bigoplus_{s\in Y} \mathcal{V}_{P_{(r,s)}}\otimes \bar{\mathcal{V}}_{P_{(r, -s)}}$.

Let us investigate how correlation functions behave in our limit. 
We write a four-point function as a sum over some spectrum in some channel. 
The sum is finite in minimal models, and becomes infinite in our limit. 
The convergence of the infinite sum depends on the behaviour of its terms as $r,s\to \infty$.
In Liouville theory, the analogous terms behave as decreasing exponentials in the total conformal dimension $\Delta+\bar\Delta$ as $\Delta+\bar\Delta\to \infty$ \cite{rs15}. 
Due to the geometric mean relation \eqref{eq:cn2cd} and the universality of conformal blocks, the same behaviour must occur in the limit of non-diagonal minimal models. 
Now, if the limiting spectrum is non-diagonal and of the type $\mathcal{S}_{X,Y}$, then the total dimension of a state is
\begin{align}
 \Delta_{(r,s)} + \Delta_{(r,-s)} = \frac{c-1}{12} + \frac12\left(r^2\beta^2 + \frac{s^2}{\beta^2}\right)\ .
\end{align}
Assuming $\Re \beta^2>0$ i.e. $\Re c<13$, this goes to infinity as $r,s\to \infty$. 
So the infinite sum over $\mathcal{S}_{X,Y}$ converges, and the finite sums that appear in minimal models tend to this infinite sum in our limit.
If however the finite sum is over diagonal states, then the situation is more subtle, because the total dimensions $2\Delta_{(r,s)}$ of diagonal degenerate states do not tend to infinity as $r,s\to \infty$. 
Our heuristic analysis of the limit of the spectrum may therefore not capture the behaviour of correlation functions, and the limiting spectrum need not necessarily be continuous or even diagonal.
Our guess is that such subtleties are absent in four-point functions of diagonal fields, whether these four fields belong to A-series minimal models, or to the diagonal sectors of D-series minimal models.
We expect that these subtleties 
occur when structure constants are not analytic as functions of momentums.
This can happen with the three-point structure constant \eqref{cdnn} due to its sign factor $f_{2,3}(P_1)$, which can be non-trivial if two fields are non-diagonal.

Therefore, we only keep the non-diagonal spectrum $\mathcal{S}_{2\mathbb{Z},\mathbb{Z}+\frac12}$ as a robust prediction from minimal models. Let us display the momentums $(P, \bar{P})$ corresponding to this spectrum (red dots) and to the spectrum of Liouville theory (thick blue line) for an arbitrarily chosen value of $\beta$:
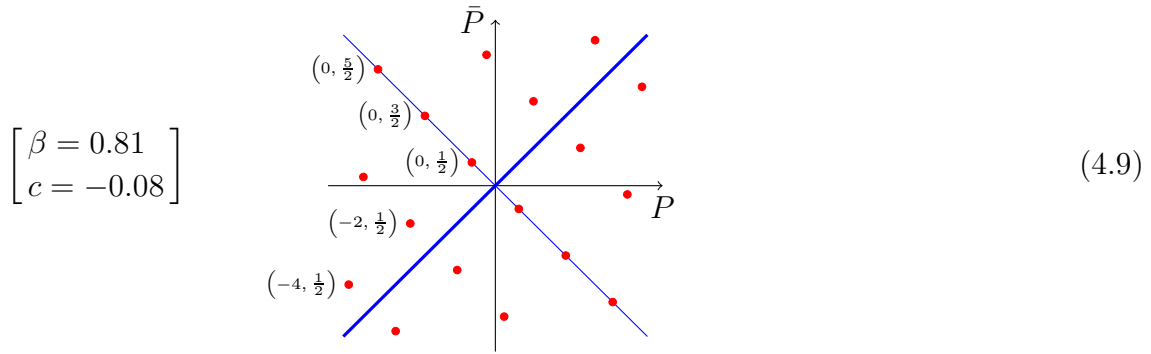
\begin{align}
\left[ \begin{array}{l}
\beta = 0.81 \\
c = -0.08
\end{array}\right]
\qquad 
\def\be{.81}
 \begin{tikzpicture}[scale = 1, baseline=(current  bounding  box.center)]
\draw [black, ->] (-2.2, 0) -- (2.2, 0) node[below]{$P$};
\draw [black, ->] (0, -2.2) -- (0, 2.2) node[left]{$\bar P$};
\draw [blue, very thick] (-2, -2) -- (2, 2);
\draw [blue, thin] (2, -2) -- (-2, 2);
\node[left] at (-2*\be/2 - 1/\be/4  , -2*\be/2 + 1/\be/4 ) {\tiny{$\left(-2, \frac{1}{2}\right)$}};
\node[left] at (-4*\be/2 - 1/\be/4  , -4*\be/2 + 1/\be/4 ) {\tiny{$\left(-4, \frac{1}{2}\right)$}};
\node[left] at (- 5/\be/4  , 5/\be/4) {\tiny{$\left(0, \frac{5}{2}\right)$}};
\node[left] at (- 3/\be/4  , 3/\be/4) {\tiny{$\left(0, \frac{3}{2}\right)$}};
\node[left] at (- 1/\be/4  , 1/\be/4) {\tiny{$\left(0, \frac{1}{2}\right)$}};
\clip (-2, 2 ) -- (-2, -2 ) -- (2, -2 ) -- (2, 2 ) -- cycle;
\foreach \x in {-4,-2,..., 4}{
  \foreach \y in {-7, -5,...,7}{
    \node[draw,circle,inner sep=1pt,fill,red] at (\x*\be/2 - \y/\be/4  , \x*\be/2 + \y/\be/4 ) {};
  }
}
 \end{tikzpicture} 
 \label{graph}
\end{align}
The momentums of $\mathcal{S}_{2\mathbb{Z},\mathbb{Z}+\frac12}$ form a lattice spanned by two vectors that point along the diagonal (thick blue line) and anti-diagonal (thin blue line). The diagonal and anti-diagonal themselves correspond to the spinless states.

A numerical bootstrap analysis in the context of the Potts model has shown that this spectrum appears in correlation functions of the type \cite{prs16}
\begin{align}
 Z_0 = \left< V^D_{P_{(0,\frac12)}} V^N_{(0,\frac12)} V^D_{P_{(0,\frac12)}} V^N_{(0,\frac12)} \right> \ .
 \label{eq:z0}
\end{align}
Our present analysis suggests that the same spectrum should appear in many more correlation functions. 
These correlation functions should conserve diagonality, as an inheritance from D-series minimal models. 
Actually, diagonality must be conserved in any theory whose non-diagonal sector is $\mathcal{S}_{2\mathbb{Z},\mathbb{Z}+\frac12}$, as a consequence of our condition \eqref{eq:sums} on three-point functions.

Let us point out that the shift equations \eqref{eq:drat} completely determine the dependence of structure constants on fields in $\mathcal{S}_{2\mathbb{Z},\mathbb{Z}+\frac12}$. This is obvious for the dependence on the first index $r$, which takes values in $2\mathbb{Z}$ while the relevant equation shifts it by $2$.
This is  less obvious for the dependence on $s$, because shifts by $2$ relate all values $s\in\mathbb{Z}+\frac12$ to two values, say $s\in\{-\frac12, \frac12\}$, rather than just one value. However, our two values of $s$ are opposite to one another, and we can use the fact that 
normalization-independent quantities are invariant under $(r,s)\to (-r,-s)$, because $V^N_{(r,s)}$ and $V^N_{(-r,-s)}$ have the same conformal dimensions.
Therefore, the solution of the shift equations is unique up to an $(r,s)$-independent factor. 
In order to compute structure constants, we can use indifferently the shift equations, or their solution.

\subsection{Numerical tests of crossing symmetry}\label{sec:nc}

We conjecture that for any central charge $c$ such that $\Re c<13$, 
for any two diagonal fields with arbitrary momentums, and any two non-diagonal fields in $\mathcal{S}_{2\mathbb{Z},\mathbb{Z}+\frac12}$, there is a crossing-symmetric four-point function 
\begin{align}
 Z = \Big< V^D_{P_1} V^N_{(r_2,s_2)} V^D_{P_3} V^N_{(r_4,s_4)} \Big> \ ,
 \label{eq:z}
\end{align}
with the spectrum $\mathcal{S}_{2\mathbb{Z},\mathbb{Z}+\frac12}$ in the $s$- and $t$-channels.
(However we know neither the $u$-channel spectrums of such four-point functions, nor the spectrums of four-point functions of the type $\left<V^NV^NV^NV^N\right>$.)

We will provide evidence for this conjecture by directly testing the crossing symmetry equation \eqref{cross} for four-point functions of the type 
\begin{align}
 Z = \Big< V^D_{P_1} V^N_{(r_2,s_2)} V^D_{P_1} V^N_{(r_2,s_2)} \Big> \ . \label{eq:z12}
\end{align}
In this case the structure constants $D_{(r,s)}$ are the same in both $s$- and $t$-channels, and are given by eq. \eqref{eq:d12}. We perform the renormalization $D_{(r,s)}\to \frac{D_{(r,s)}}{D_{(0,\frac12)}}$, in order to have
\begin{align}
D_{(0,\frac{1}{2})} = 1\, .
\end{align}
The four-point structure constants were computed iteratively by using \eqref{eq:drat}, taking the ground-state structure constant $D_{(0,\frac{1}{2})}$ as a starting point. Conformal blocks were computed using Zamolodchikov's recursion formula, see appendix A.2 in \cite{rs15} for more details. 
The Jupyter notebooks used to perform these computations are available on \href{https://github.com/ribault/bootstrap-2d-Python}{GitHub}, along with further examples.

Before testing crossing symmetry of generic four-point functions
let us verify that our analytic four-point structure constants \eqref{eq:d12} agree with the numerically determined structure constants in \cite{prs16}, for the four-point function $Z_0$ \eqref{eq:z0} in the case $c=0$ that is relevant for critical percolation. The following table shows the values of the first $9$ four-point structure constants calculated with each method. The coefficient of variation $c_{(r,s)}$ is an estimate of the precision of the numerical bootstrap determination of the corresponding structure constant. The last column shows the relative difference between both calculations. 
\begin{align}
\begin{array}{|c|cr|c|c|} \hline 
 & \text{Numerical bootstrap} & & \text{Analytic bootstrap} & \text{Relative} \\
(r,s) & D_{(r,s)} & c_{(r,s)} & D_{(r,s)} & \text{differences} \\ \hline 
\left ( 0,  \frac{1}{2}\right )& 1 & 0& 1& 0 \\ 
\left ( 2,  \frac{1}{2}\right )& \phantom{+}0.0385548051& 2.4 \times 10^{-9}& \phantom{+}0.0385548051& 8.7 \times 10^{-10} \\ 
\left ( 0,  \frac{3}{2}\right )& -0.0212806511& 7.6 \times 10^{-9}& -0.0212806510& 2.9 \times 10^{-9} \\ 
\left ( 2,  \frac{3}{2}\right )& \phantom{+}0.0004525024& 2.2 \times 10^{-8}& \phantom{+}0.0004525024& 1.7 \times 10^{-9} \\ 
\left ( 0,  \frac{5}{2}\right )& -3.5638 \cdot 10^{-5}& 4.4 \times 10^{-7}& -3.5638 \cdot 10^{-5}& 3.7 \times 10^{-8} \\ 
\left ( 4,  \frac{1}{2}\right )& -2.9746 \cdot 10^{-6}& 2.4 \times 10^{-6}& -2.9746 \cdot 10^{-6}& 1.2 \times 10^{-6} \\ 
\left ( 2,  \frac{5}{2}\right )& \phantom{+}8.4077 \cdot 10^{-7}& 1.3 \times 10^{-5}& \phantom{+}8.4078 \cdot 10^{-7}& 6.3 \times 10^{-6} \\ 
\left ( 4,  \frac{3}{2}\right )& -4.4131 \cdot 10^{-8}& 1.6 \times 10^{-4}& -4.4135 \cdot 10^{-8}& 8.8 \times 10^{-5} \\ 
\left ( 0,  \frac{7}{2}\right )& \phantom{+}1.5064 \cdot 10^{-10}& 9.3 \times 10^{-3} & \phantom{+}1.5174 \cdot 10^{-10}& 7.3 \times 10^{-3} \\ 
\hline
\end{array}
\end{align}
We note that not only the absolute values, but also the signs of the structure constants agree, a result that could not be deduced from the geometric mean formula \eqref{eq:csqrt}. While the precision of the numerical bootstrap calculations decreases as $s$-channel conformal dimensions increase, our analytic results (coming from \eqref{eq:drat}) do not have this problem. This explains the increase in relative differences between both calculations, and the fact that the estimated numerical uncertainty $c_{(r,s)}$ is comparable to these differences. This supports the idea that the analytic results are indeed exact.

Let us now test crossing symmetry of four-point functions computed from our analytic structure constants. For a number of choices of the parameters $c, \Delta_1$ and $(r_2,s_2)$, we will display the values of the four-point function $Z$ \eqref{eq:z12} (or its real part when it is complex) computed from the $s$- and $t$-channels for four values of the cross-ratio $z$, and the relative difference $\left|2\frac{Z^{(s)}-Z^{(s)}}{Z^{(s)}+Z^{(t)}}\right|$ between the two channels. Our first choice of parameters corresponds again to the four-point function $Z_0$ at $c=0$:
\begin{align}
\left[\begin{array}{l}
c = 0 \\
\Delta_1 =\Delta_{(0,\tfrac{1}{2})}\\
(r_2,s_2) =(0,\tfrac{1}{2})\\
\end{array}\right] \ \ \rightarrow \ \ 
\begin{array}{|c|c|c|} \hline 
 z & Z_{0}(z) & \text{Difference} \\ \hline
 0.01 & \begin{array}{l}  
s: 0.420743288653023 \\
 t: 0.420743500577090 
\end{array} & 5 \times 10^{-7} \\ \hline
 0.03 & \begin{array}{l}  
s: 0.514703102283562 \\
 t: 0.514703104911165 
\end{array} & 5.1 \times 10^{-9} \\ \hline
 0.1 & \begin{array}{l}  
s: 0.635102793381169 \\
 t: 0.635102793359283 
\end{array} & 3.4 \times 10^{-11} \\ \hline
 0.2 & \begin{array}{l}  
s: 0.706457914575874 \\
 t: 0.706457914575509 
\end{array} & 5.2 \times 10^{-13} \\ \hline
 0.4 & \begin{array}{l}  
s: 0.761209621824938 \\
 t: 0.761209621824937 
\end{array} & 8.8 \times 10^{-16} \\ \hline
\end{array}
\end{align}
\begin{align}
\left[\begin{array}{l}
c = 0.7513 \\
\Delta_1 =0.0731 \\
(r_2,s_2) =(0,\tfrac{3}{2})\\
\end{array}\right] \ \ \rightarrow \ \ 
\begin{array}{|c|c|c|} \hline 
 z & Z(z) & \text{Difference} \\ \hline
 0.01 & \begin{array}{l}  
s: 0.458407080230741 \\
 t: 0.458407080231849 
\end{array} & 2.4 \times 10^{-12} \\ \hline
 0.03 & \begin{array}{l}  
s: 0.453858548437590 \\
 t: 0.453858548437598 
\end{array} & 1.8 \times 10^{-14} \\ \hline
 0.1 & \begin{array}{l}  
s: 0.166547888308668 \\
 t: 0.166547888308669 
\end{array} & 8 \times 10^{-15} \\ \hline
 0.2 & \begin{array}{l}  
s: -0.350128460299570 \\
 t: -0.350128460299570 
\end{array} & 1.4 \times 10^{-15} \\ \hline
 0.4 & \begin{array}{l}  
s: -1.085635947272218 \\
 t: -1.085635947272219 
\end{array} & 8.2 \times 10^{-16} \\ \hline
\end{array}
\end{align}
\begin{align}
\left[\begin{array}{l}
c = 0.7513 \\
\Delta_1 =0.0731 \\
(r_2,s_2) =(2,\tfrac{1}{2})\\
\end{array}\right] \ \ \rightarrow \ \ 
\begin{array}{|c|c|c|} \hline 
 z & Z(z) & \text{Difference} \\ \hline
 0.01 & \begin{array}{l}  
s: 0.454365494340930 \\
 t: 0.454365494931425 
\end{array} & 1.3 \times 10^{-9} \\ \hline
 0.03 & \begin{array}{l}  
s: 0.488568249965185 \\
 t: 0.488568249967384 
\end{array} & 4.5 \times 10^{-12} \\ \hline
 0.1 & \begin{array}{l}  
s: 0.603030367288685 \\
 t: 0.603030367288691 
\end{array} & 9.9 \times 10^{-15} \\ \hline
 0.2 & \begin{array}{l}  
s: 0.968890687817652 \\
 t: 0.968890687817658 
\end{array} & 6.6 \times 10^{-15} \\ \hline
 0.4 & \begin{array}{l}  
s: 1.720792857730145 \\
 t: 1.720792857730149 
\end{array} & 1.9 \times 10^{-15} \\ \hline
\end{array}
\end{align}
\begin{align}
\left[\begin{array}{l}
c = 4.72+0.12i \\
\Delta_1 = 0.231+0.1432i\\
(r_2,s_2) =(0,\tfrac{1}{2})\\
\end{array}\right] \ \ \rightarrow \ \ 
\begin{array}{|c|c|c|} \hline 
 z & \Re{Z}(z) & \text{Difference} \\ \hline
 0.01 & \begin{array}{l}  
s: 0.616734633551431 \\
 t: 0.616734633551427 
\end{array} & 1.3 \times 10^{-14} \\ \hline
 0.03 & \begin{array}{l}  
s: 0.712923810169360 \\
 t: 0.712923810169357 
\end{array} & 9.9 \times 10^{-15} \\ \hline
 0.1 & \begin{array}{l}  
s: 0.784039104794377 \\
 t: 0.784039104794372 
\end{array} & 6.5 \times 10^{-15} \\ \hline
 0.2 & \begin{array}{l}  
s: 0.772087133344852 \\
 t: 0.772087133344848 
\end{array} & 5.3 \times 10^{-15} \\ \hline
 0.4 & \begin{array}{l}  
s: 0.724037027055028 \\
 t: 0.724037027055025 
\end{array} & 3.9 \times 10^{-15} \\ \hline
\end{array}
\end{align}
\begin{align}
\left[\begin{array}{l}
c = 4.72+0.12i \\
\Delta_1 = 0.231+0.1432i\\
(r_2,s_2) =(2,\tfrac{3}{2})\\
\end{array}\right] \ \ \rightarrow \ \ 
\begin{array}{|c|c|c|} \hline 
 z & \Re{Z}(z) & \text{Difference} \\ \hline
 0.01 & \begin{array}{l}  
s: 0.642349222078378 \\
 t: 0.642607584797052 
\end{array} & 4.0 \times 10^{-4} \\ \hline
 0.03 & \begin{array}{l}  
s: -0.111756545727778 \\
 t: -0.111755334852274 
\end{array} & 5.5 \times 10^{-6} \\ \hline
 0.1 & \begin{array}{l}  
s: -2.187839195998956 \\
 t: -2.187839195940312 
\end{array} & 5.4 \times 10^{-11} \\ \hline
 0.2 & \begin{array}{l}  
s: -4.295291665306662 \\
 t: -4.295291665306607 
\end{array} & 1.9 \times 10^{-14} \\ \hline
 0.4 & \begin{array}{l}  
s: -13.166871727284267 \\
 t: -13.166871727284297 
\end{array} & 7.6 \times 10^{-15} \\ \hline
\end{array}
\end{align}
We obtain larger differences for some values of the parameters, because of numerical truncations in calculations of conformal blocks. Of course, differences increase as $z\to 0$, where the $t$-channel expansion diverges. Moreover, values of $c\in (-\infty, 1)$ suffer from the proximity of poles at the minimal model values $c=c_{p,q}$. And larger values of $\Delta_1,\Delta_{(r_2,s_2)}$ lead to larger errors. 

Overall, we find strong evidence that crossing symmetry is satisfied. In particular, the observed discrepancies between $s$-channel and $t$-channel results are controlled by the truncation that we use for computing conformal blocks using Zamolodchikov's recursion formula, and we find no hint that we should add more states in our spectrum. 
This supports our claim that there exist
consistent non-rational conformal field theories whose non-diagonal spectrum is $\mathbb{S}_{2\mathbb{Z}, \mathbb{Z}+\frac{1}{2}}$, and whose structure constants satisfy the analytic bootstrap equations of Section \ref{sec:acb}.

\subsection{Case of the Ashkin--Teller model}\label{sec:atm}

The Ashkin--Teller model provides an example of a four-point function that is known analytically, and does not conserve diagonality. The model has the central charge $c=1$, and it has a four-point function of the type \cite{zam85b}
\begin{align}
 \left< V^D_{P_{(0,\frac12)}}V^D_{P_{(0,\frac12)}}V^N_{(0,\frac12)}V^N_{(0,\frac12)}\right> = \sum_{r\in 2\mathbb{Z}}\sum_{s\in\mathbb{Z}} D_{(r,s)}  \mathcal{F}^{(s)}_{\Delta_{(r,s)}} \bar{\mathcal{F}}^{(s)}_{\Delta_{(r,-s)}}\ , \label{eq:4ptat}
\end{align}
with the four-point structure constants 
\begin{align}
 D_{(r,s)} = (-1)^{\frac{r}{2}} 16^{-\frac12 r^2 -\frac12 s^2}\ .
 \label{eq:drs}
\end{align}
Let us see whether this obeys our shift equations \eqref{eq:drat}. We first compute the relevant ratios $\rho$. In the case $c=1$, eq. \eqref{OPEcg} reduces to 
\begin{align}
 \rho(V) = \frac{\sin (2\pi \sigma \bar P)}{\sin(2\pi  P)} \quad \implies \quad \rho(V^N_{(r,s)}) = (-1)^{2s}\ .
\end{align}
Then let us evaluate eq. \eqref{eq:r} with $\beta=1$ and $P_2=\bar P_2=P_3=\bar P_3 = \frac{1}{4}$. Using the Gamma function's duplication formula, and more specifically its consequence
\begin{align}
 \prod_{\pm, \pm}\Gamma(\tfrac12 \pm \tfrac14\pm\tfrac14 +x) = 2^{2-4x}\pi\Gamma(2x)\Gamma(2x+1)\ ,
\end{align}
we find 
\begin{align}
 \rho(V_1|V_2,V_3) = (-1)^{2s_2} 2^{-4(P_1+\sigma_1\bar P_1)} \frac{\sin(2\pi \sigma_1\bar P_1)}{\sin(2\pi  P_1)}\ .
\end{align}
We are interested in two cases of this ratio, where the fields $V_2,V_3$ are either diagonal (thus $s_2=0$) or non-diagonal (with $s_2=\frac12$). In these two cases, we find 
\begin{align}
 \rho\left(V^N_{(r,s)}\middle|V^N_{(0,\frac12)},V^N_{(0,\frac12)}\right) = -(-1)^{2s} 16^{-r} \quad , \quad \rho\Big(V^N_{(r,s)}\Big|V^D_{P_{(0,\frac12)}},V^D_{P_{(0,\frac12)}}\Big) = (-1)^{2s} 16^{-r}\ .
\end{align}
For non-diagonal fields with $s\in\mathbb{Z}$, the shift equation therefore reduces to 
\begin{align}
 \frac{D_{(r+1,s)}}{D_{(r-1,s)}} = -16^{-2r}\ ,
\end{align}
and this is indeed obeyed by the four-point structure constant \eqref{eq:drs}. By a similar analysis, we would find the second shift equation $\frac{D_{(r,s+1)}}{D_{(r,s-1)}} = 16^{-2s}$, where we no longer have a minus sign because the non-diagonal field $V^N_{(0,\frac12)}$ has an integer first index. 

The four-point function \eqref{eq:4ptat} is a sum over the spectrum $\mathcal{S}_{2\mathbb{Z},\mathbb{Z}}$. In this case, our shift equations relate all the four-point structure constants to $D_{(0,0)}$ and $D_{(0,1)}$, but do not relate these two numbers to one another. Nevertheless, their compatibility with the known structure constants is a non-trivial test of our ideas and calculations. And this case illustrates the difference between diagonal and non-diagonal fields with identical conformal dimensions, namely $V^D_{P_{(0,\frac12)}}$ and $V^N_{(0,\frac12)}$.

\section{Conclusion}

Our results suggest that for any central charge $c$ such that $\Re c<13$, there exists a non-rational conformal field theory, whose spectrum has the non-diagonal sector $\mathcal{S}_{2\mathbb{Z},\mathbb{Z}+\frac12}$.
The three-point structure constants in that theory are given by our analytic formulas. 
For $c< 1$, that theory is a limit of D-series minimal models. 
Some particular four-point functions in that theory describe connectivities of clusters in the critical Potts model, at least as very good approximations \cite{prs16}.

The three-point structure constants that we have determined should be valid not only in that theory, but actually in any CFT that obeys our assumptions, starting with the existence of two independent degenerate fields. 
A second theory of the same type can be obtained by $\beta \to \frac{1}{\beta}$, and its non-diagonal sector is $\mathcal{S}_{\mathbb{Z}+\frac12,2\mathbb{Z}}$.
But not all interesting CFTs obey our assumptions. For example, the sigma models of \cite{rs01} have non-diagonal fields whose indices $r$ take fractional values, rather than our half-integer values. 
This makes it implausible that the degenerate field $V_{\langle 2, 1\rangle}$ exists, as fusion with this field would shift $r$ by integers. One might still be tempted to use the formula \eqref{eq:csqrt} for the structure constants, but this would leave us with an undetermined sign and would most probably be wrong. 
In such a case, a more promising approach would be to renounce analytic formulas, and determine structure constants using the numerical bootstrap method of \cite{prs16}.


\acknowledgments{
We are grateful to Benoît Estienne, Yacine Ikhlef and 
Raoul Santachiara for stimulating discussions and comments on this text. We also thank to the two anonymous JHEP referees for their suggestions.
}


\appendix

\section{Universality of the geometric mean relation}\label{app}

Here we show that under rather general conditions, there must be a geometric mean relation of the type of \eqref{eq:csqrt} between structure constants of diagonal and non-diagonal conformal field theories. For some more detail, see \cite{rib18}.

\subsection{Mathematical statement}

Let $D^+$ and $D^-$ be two meromorphic differential operators of order $n$ on the Riemann sphere. 
Let a non-diagonal solution of $(D^+,D^-)$ be a single-valued function $f$ such that $D^+f = \bar{D}^- f=0$, where $\bar D^-$ is obtained from $D^-$ by $z\to\bar z,\partial_z \to \partial_{\bar z}$. Let a diagonal solution of $D^+$ be a single-valued function $f$ such that $D^+f =\bar D^+f=0$.

We assume that $D^+$ and $D^-$ have singularities 
at two points $0$ and $1$. Let $(\mathcal F^\epsilon_i)$ and $(\mathcal{G}^\epsilon_i)$ be bases of solutions of $D^\epsilon f=0$ that diagonalize the monodromies around $0$ and $1$ respectively. 
In the case of $(\mathcal F^+_i)$ this means 
\begin{align}
 D^+ \mathcal{F}^+_i = 0 \quad , \quad \mathcal{F}^+_i \left(e^{2\pi i}z\right) = \lambda_i \mathcal{F}^+_i(z)\ .
\end{align}
We further assume that our bases are such that 
\begin{align}
 \forall \epsilon,\bar{\epsilon}\in\{+,-\}\, , \quad \left\{ 
 \begin{array}{l}
  \mathcal{F}^\epsilon_i(z) \mathcal{F}^{\bar\epsilon}_j(\bar z) \ \text{has no monodromy around } z=0 \ \ \iff \ \ i=j\ ,
  \\
  \mathcal{G}^\epsilon_i(z) \mathcal{G}^{\bar\epsilon}_j(\bar z) \ \text{has no monodromy around } z=1 \ \ \iff \ \ i=j\ .
 \end{array}\right.
\label{tmo}
\end{align}
For $\epsilon \neq \bar{\epsilon}$ this is a rather strong assumption, which implies that the operators $D^+$ and $D^-$ are closely related to one another. This assumption implies that a non-diagonal solution $f^0$ has expressions of the form
\begin{align}
 f^0(z,\bar z) = \sum_{i=1}^n c^0_i \mathcal{F}_i^+(z) \mathcal{F}_i^-(\bar z) = \sum_{i=1}^n d^0_i \mathcal{G}^+_i(z) \mathcal{G}_i^-(\bar z)\ ,
 \label{fz}
\end{align}
for some structure constants $(c^0_i)$ and $(d^0_i)$. Similarly, a diagonal solution $f^\epsilon$ of $D^\epsilon$ has expressions of the form
\begin{align}
 f^\epsilon(z,\bar z) = \sum_{i=1}^n c^\epsilon_i \mathcal{F}_i^\epsilon(z) \mathcal{F}_i^\epsilon(\bar z) = \sum_{i=1}^n d^\epsilon_i \mathcal{G}^\epsilon_i(z) \mathcal{G}_i^\epsilon(\bar z)\ .
 \label{fe}
\end{align}
We now claim that 
\begin{quote}
 if $D^+$ and $D^-$ have diagonal solutions, and if moreover $(D^+,D^-)$ has a non-diagonal solution, then the non-diagonal structure constants are geometric means of the diagonal structure constants, 
 \begin{align}
   (c^0_i)^2 \propto c^+_ic^-_i\ , 
   \label{ccc}
 \end{align}
 where $\propto$ means equality up to an $i$-independent prefactor.
\end{quote}
The proof of this statement is simple bordering on the trivial. 
We introduce the size $n$ matrices $M^\epsilon$ such that 
\begin{align}
 \mathcal{F}^\epsilon_i = \sum_{j=1}^n M^\epsilon_{i,j} \mathcal{G}^\epsilon_j \ .
\end{align}
Inserting this change of bases in eq. \eqref{fe}, we must have 
\begin{align}
 j\neq k \implies \sum_{i=1}^n c^\epsilon_i M_{i,j}^\epsilon M_{i,k}^\epsilon = 0\ .
\end{align}
For a given $\epsilon$, this is a system of $\frac{n(n-1)}{2}$ linear equations for $n$ unknowns $c^\epsilon_i$. One way to write the solution is 
\begin{align}
 c^\epsilon_i \propto (-1)^i\det_{\substack{ i'\neq i \\ j \neq 1}} \left( M^\epsilon_{i',1}M^\epsilon_{i',j} \right) 
 = (-1)^i \left(\prod_{i'\neq i} M^\epsilon_{i',1}\right)  \det_{\substack{ i'\neq i \\ j \neq 1}} \left( M^\epsilon_{i',j} \right)\ .
\end{align}
Similarly, inserting the change of bases in the expression \eqref{fz} of a non-diagonal solution, we find 
\begin{align}
 j\neq k \implies \sum_{i=1}^n c^0_i M_{i,j}^+ M_{i,k}^- = 0\ .
\end{align}
We will write two expressions for the solution of this linear equations,
\begin{align}
 c^0_i &\propto (-1)^i \det_{\substack{ i'\neq i \\ j \neq 1}} \left(  M^-_{i',1}M^+_{i',j}\right) = (-1)^i \left(\prod_{i'\neq i} M^-_{i',1}\right)  \det_{\substack{ i'\neq i \\ j \neq 1}} \left( M^+_{i',j} \right)\ ,
 \\
       &\propto (-1)^i \det_{\substack{ i'\neq i \\ j \neq 1}} \left(  M^+_{i',1}M^-_{i',j}\right) = (-1)^i \left(\prod_{i'\neq i} M^+_{i',1}\right)  \det_{\substack{ i'\neq i \\ j \neq 1}} \left( M^-_{i',j} \right)\ .
\end{align}
Writing $(c^0_i)^2$ as the product of the above two expressions, we obtain eq. \eqref{ccc}.

The difficult problem is actually to study the conditions on the matrices $M^\epsilon$ for diagonal and non-diagonal solutions to exist. It appears that the existence of non-diagonal solutions is in general equivalent to 
\begin{align}
 \forall i,j, \ \ M^+_{ij} \left((M^+)^{-1}\right)_{ji} = M^-_{ij} \left((M^-)^{-1}\right)_{ji}\ ,
\end{align}
but this equation is not easy to solve.

\subsection{Conformal field theory applications}

In our case, the differential operators $D^+$ and $D^-$ are respectively left- and right-moving BPZ equations of the order $n=2$. The solutions $\mathcal F^\epsilon_i$ and $\mathcal G^\epsilon_i$ are respectively $s$- and $t$-channel conformal blocks, and the matrices $M^+$ and $M^-$ are respectively the left- and right-moving fusing matrices. The structure constants $c^0_i$ and $d^0_i$ are respectively $s$- and $t$-channel four-point structure constants for a four-point function of one degenerate field, and three fields that may be non-diagonal.

Then our mathematical statement is the geometric mean relation \eqref{rhrh} for such degenerate four-point structure constants. But we have seen in Section \ref{ssec:OPE} how these degenerate structure constants determine more general three- and four-point structure constants via shift equations. 
This is why the geometric mean relation holds in four-point functions that do not involve degenerate fields, although such four-point functions violate our assumptions: they are combinations of infinitely many conformal blocks, do not obey differential equations, and violate the assumption \eqref{tmo} because for example both $\mathcal{F}^{(s)}_{\Delta_{(r,s)}}(z)\bar{\mathcal{F}}^{(s)}_{\Delta_{(r,s)}}(\bar z)$ and $\mathcal{F}^{(s)}_{\Delta_{(r,s)}}(z)\bar{\mathcal{F}}^{(s)}_{\Delta_{(r,-s)}}(\bar z)$ are single-valued if $rs\in\mathbb{Z}$.

Our derivation of the geometric mean relation implies that it holds for four-point functions that obey BPZ equations of any order. It can also hold in CFTs with W-algebras, provided our assumptions are obeyed. The assumption \eqref{tmo} may be difficult to satisfy for rational central charges, because conformal dimensions of two fields can easily differ by integers. It looks easier to satisfy when the central charge is generic.


\begin{thebibliography}{10}
\expandafter\ifx\csname url\endcsname\relax
  \def\url#1{\texttt{#1}}\fi
\expandafter\ifx\csname urlprefix\endcsname\relax\def\urlprefix{URL }\fi
\providecommand{\eprint}[2][]{\url{#2}}

\bibitem{rib16}
\href{https://inspirehep.net/record/1489097/files/arXiv:1609.09523.pdf}{S.~Ribault}
  (2016 review) [1609.09523]\\ {\em {Minimal lectures on two-dimensional
  conformal field theory}\/}

\bibitem{ei15}
\href{http://arxiv.org/abs/1505.00585}{B.~Estienne, Y.~Ikhlef} (2015)
  [1505.00585]\\ {\em {Correlation functions in loop models}\/}

\bibitem{Ribault:2014hia}
\href{http://arxiv.org/abs/1406.4290}{S.~Ribault} (2014 review) [1406.4290]\\
  {\em {Conformal field theory on the plane}\/}

\bibitem{hj13}
\href{http://arxiv.org/abs/1312.4520}{L.~Hadasz, Z.~Jaskólski} (2014)
  [1312.4520]\\ {\em {Super-Liouville - Double Liouville correspondence}\/}

\bibitem{rs15}
\href{http://arxiv.org/abs/1503.02067}{S.~Ribault, R.~Santachiara} (2015)
  [1503.02067]\\ {\em {Liouville theory with a central charge less than one}\/}

\bibitem{zam85b}
{\relax Al}.~B. Zamolodchikov (1986)\\ {\em Two-dimensional conformal symmetry
  and critical four-spin correlation functions in the Ashkin-Teller model\/}

\bibitem{fms97}
P.~Di~Francesco, P.~Mathieu, D.~Senechal (1997 book)\\ {\em Conformal field
  theory\/}

\bibitem{run99}
\href{http://arxiv.org/abs/hep-th/9908046}{I.~Runkel} (2000) [hep-th/9908046]\\
  {\em {Structure constants for the D series Virasoro minimal models}\/}

\bibitem{prs16}
\href{http://arxiv.org/abs/1607.07224}{M.~Picco, S.~Ribault, R.~Santachiara}
  (2016) [1607.07224]\\ {\em {A conformal bootstrap approach to critical
  percolation in two dimensions}\/}

\bibitem{rs01}
\href{http://arxiv.org/abs/hep-th/0106124}{N.~Read, H.~Saleur} (2001)
  [hep-th/0106124]\\ {\em {Exact spectra of conformal supersymmetric nonlinear
  sigma models in two-dimensions}\/}

\bibitem{rib18}
\href{http://researchpracticesandtools.blogspot.fr/2018/01/on-single-valued-solutions-of.html}{S.~Ribault}
  (2018 blog post)\\ {\em On single-valued solutions of differential
  equations\/}

\end{thebibliography}
\end{document}